\documentclass[a4paper,fleqn]{cas-dc}

\usepackage[authoryear]{natbib}
\usepackage{dirtytalk}
\usepackage{amsmath,bm,physics2,derivative, cancel,siunitx}
\usepackage{subfigure}
\usephysicsmodule{ab.legacy}

\usepackage[capitalise, nameinlink]{cleveref}

\renewcommand{\vec}[1]{\bm{#1}}

\DeclareMathAlphabet{\MathSfBI}{OT1}{\sfdefault}{bx}{sl}
\DeclareMathVersion{sfBLetters}
\SetSymbolFont{letters}{sfBLetters}{OML}{ntxsfmi}{b}{it}
\newcommand{\MathSfBILow}[1]{%
	\text{\mathversion{sfBLetters}$\m@th#1$}%
}
\DeclareRobustCommand{\TnsrBoldImpl}[1]{%
	\begingroup
	\ifcat\noexpand #1\relax
		\edef\greek@test{\detokenize{#1}}%
		\edef\greek@test{\expandafter\@cdr\greek@test\@nil}%
		\edef\greek@test{\expandafter\@car\greek@test\@nil}%
		\edef\x{\the\lccode\expandafter`\greek@test}%
		\edef\y{\number\expandafter`\greek@test}%
		\ifnum\x=\y\relax
			\MathSfBILow{#1}%
		\else
			\MathSfBI{#1}%
		\fi
	\else
		\MathSfBI{#1}%
	\fi
	\endgroup
}

\DeclareMathAlphabet{\MathSfI}{OT1}{\sfdefault}{m}{sl}
\DeclareMathVersion{sfLetters}
\SetSymbolFont{letters}{sfLetters}{OML}{ntxsfmi}{m}{it}
\newcommand{\MathSfILow}[1]{%
	\text{\mathversion{sfLetters}$\m@th#1$}%
}
\DeclareRobustCommand{\TnsrImpl}[1]{%
	\begingroup
	\ifcat\noexpand #1\relax
		\edef\greek@test{\detokenize{#1}}%
		\edef\greek@test{\expandafter\@cdr\greek@test\@nil}%
		\edef\greek@test{\expandafter\@car\greek@test\@nil}%
		\edef\x{\the\lccode\expandafter`\greek@test}%
		\edef\y{\number\expandafter`\greek@test}%
		\ifnum\x=\y\relax
			\MathSfILow{#1}%
		\else
			\MathSfI{#1}%
		\fi
	\else
		\MathSfI{#1}%
	\fi
	\endgroup
}

\makeatother

\DeclareMathAlphabet{\mathsfit}{\encodingdefault}{\sfdefault}{m}{sl}
\SetMathAlphabet{\mathsfit}{bold}{\encodingdefault}{\sfdefault}{bx}{sl}

\NewDocumentCommand{\tnsr}{sm}{%
	\IfBooleanTF{#1}{%
		\TnsrImpl{#2}%
	}{%
		\TnsrBoldImpl{#2}%
	}
}

\NewDocumentCommand{\per}{m}{%
	\breve{#1}%
}

\NewDocumentCommand{\tavg}{sm}{%
	\IfBooleanTF{#1}{{#2}^{\prime}}{\overline{#2}}
}

\NewDocumentCommand{\fil}{sm}{%
	\IfBooleanTF{#1}{{#2}^{\prime\prime}}{\widehat{#2}}
}

\NewDocumentCommand{\va}{t_sm}{%
	\IfBooleanTF{#1}
	{%
		\IfBooleanTF{#2}
		{\errmessage{No fluctuation for superficial avg!}}
		{{\left\langle{#3}\right\rangle}_{\!s}}%
	}%
	{%
		\IfBooleanTF{#2}
		{\tilde{#3}}
		{{\left\langle{#3}\right\rangle}}%
	}%
}

\NewDocumentCommand{\da}{m}{\va{\tavg{#1}}}

\newcommand{\Ma}{\ensuremath{\mathrm{Ma}}}
\renewcommand{\Pr}{\ensuremath{\mathrm{Pr}}}

\renewcommand{\Re}{\ensuremath{\mathrm{Re}}}
\newcommand{\sgs}{\ensuremath{\mathrm{sgs}}}

\newcommand{\LES}{Large Eddy Simulation (LES)}
\newcommand{\DNS}{Direct Numerical Simulation (DNS)}

\begin{document}
\let\WriteBookmarks\relax
\def\floatpagepagefraction{1}
\def\textpagefraction{.001}

\shorttitle{Influence of Prandtl number on heat transfer over a permeable wall}

\shortauthors{W. Sadowski, H. Demir and F. {di} Mare}

\title[mode = title]{Influence of Prandtl number on heat transfer over a permeable wall}

\tnotemark[1]
\tnotetext[1]{This work was funded by the Deutsche Forschungsgemeinschaft (DFG,
	German Research Foundation) – Project-ID 422037413 - TRR 287. The authors also
	gratefully acknowledge the Gauss Centre for Supercomputing e.V. (https://
	www.gauss-centre.eu) for providing computing time on the GCS Supercomputer
	Super-MUC–NG at the Leibniz Supercomputing Centre (https://www.lrz.de).}

\author[1]{Wojciech Sadowski}[orcid=0000-0002-1090-9109]
\ead{wojciech.sadowski@rub.de}
\credit{
	Conceptualization,
	Methodology,
	Software,
	Validation,
	Formal analysis,
	Investigation,
	Resources (supporting),
	Data Curation,
	Writing - Original Draft,
	Writing - Review \& Editing,
	Visualization,
	Funding acquisition (supporting)%
}

\author[1]{Hakan Demir}[orcid=0009-0006-3929-8352]
\ead{h.demir@rub.de}
\credit{
	Formal analysis (supporting),
	Resources (supporting),
	Data Curation (supporting),
	Writing - Original Draft (supporting),
	Writing - Review \& Editing (supporting)
}

\affiliation[1]{
organization={Chair of Thermal Turbomachines and Aeroengines, Ruhr University Bochum},
addressline={Universit\"{a}tsstra\ss{}e 150},
city={Bochum},
citysep={}, 
postcode={D-44801},
country={Germany}
}

\author[1]{Francesca di Mare}
\cormark[1]
\ead{francesca.dimare@rub.de}
\credit{
	Resources,
	Writing - Review \& Editing (supporting),
	Supervision,
	Project administration,
	Funding acquisition%
}

\cortext[1]{Corresponding author}

\begin{abstract}
	The work considers a fully turbulent flow with heat transfer in a channel
	half-filled with an array of cubes based on the work of \citet{breugem2005}
	and \citet{chandesris2013}, at $\Re_\mathrm{bulk} = 5485$ and three different
	Prandtl numbers, $\Pr = 0.71,
		0.1, 0.05$. The temperature is modelled as a passive scalar and two different
	boundary condition configurations are simulated. The influence of the
	Prandtl number on the mean temperature, its variance and the terms of
	the temperature budget is highlighted, including the analysis of the
	distribution and relative importance of the turbulent heat transfer,
	molecular diffusion, tortuosity and Brinkman terms near the porous-fluid
	interface. The latter two has been found to be insignificant for the
	highest $\Pr$. A set of terms, typically neglected during the upscaling
	procedure (related to the Taylor expansion of the filtered variables), is
	analysed for the first time for the turbulent heat transfer at the porous-fluid interface, and
	are found to be significant at low $\Pr$. The upscaled fields are
	evaluated with three different kernels forming cellular average, linear
	(i.e., tent kernel), quadratic and cubic, and the influence of the chosen
	filter is additionally studied.
\end{abstract}


%


\begin{keywords}
	Porous-fluid interface\sep Turbulent heat transfer\sep Double-averaged equations
\end{keywords}

\maketitle

\section{Introduction}

\label{sec:introduction}

Flows over permeable walls, or objects which can be modelled as such (for
example, trees, urban canopies, rocky riverbeds, packed beds) appear in
environmental, process and aerospace applications
\citep{Nepf2012,lowe2008,Cerminara2019}. The elements of the solid phase (e.g.,
the particles constituting the packing in the packed bed) can than be modelled
in aggregate as a porous medium and its surface acts as a permeable, rough
wall. It couples the porous flow in between the elements of the solid phase and
unobstructed flow outside, hence it is called the porous-fluid interface (PFI).
This region is important for packed-bed reactors \citep{collazo2012},
transpiration cooling in turbines and reentry vehicles \citep{Hermann2021},
ablative heat shields \citep{mansour2024}, and hyporheic exchange in rivers and
coastal ecosystems \citep{lowe2008,Finnigan2000,Shen2020}.

At the PFI, fluid develops a finite slip velocity due to the permeability of
the porous medium. Moreover, \DNS{} studies show that wall permeability
intensifies turbulence \citep{suga2010,breugem2006,kuwata2016}, raises Reynolds
stresses above the values found over impermeable walls of equivalent roughness
\citep{kuwata2016}, and lowers the transition Reynolds number
\citep{rosti2018,suga2010}. Skin friction over the porous region is also higher
than over an impermeable wall \citep{breugem2005,kuwata2016}. The mean velocity
profile is inflectional in the vicinity of the interface, suggesting that the
Kelvin--Helmholtz rollers dominate momentum and scalar exchange between the
freeboard and the pore space \citep{breugem2006,chandesris2013}.
\citet{kuwata2019} revealed a presence of two coexisting perturbation modes
in the turbulence near the permeable wall, streamwise mode associated with the
Kelvin--Helmholtz instability and spanwise one, with their relative strength
modulated by the permeability Reynolds number.

The initial picture of the turbulent heat transfer has been provided by the
work of \citet{chandesris2013} focusing on the thermal transport at $\Pr = 0.1$
in the geometry of \citet{breugem2005}, using three different heating
arrangements. They reported that Kelvin–Helmholtz vortices enhance turbulent
heat transfer above the porous region and that the pressure waves generated by
these structures penetrate deep into the porous layer. Later studies have
broadened the thermal picture, however, without systematically extending the
parameter range of \citet{chandesris2013}. \citet{kuwata2020}, in a
lattice-Boltzmann DNS of conjugate heat transfer in a porous-walled duct, found
that mean-velocity dispersion contributes as strongly as turbulent fluctuations
to energy transport inside the porous layer. \citet{nishiyama2020a} considered
turbulent flow in a channel with porous layer at the bottom, assuming heat
transfer at $\Pr=0.71$. Various geometries of the porous medium were
considered, and for each, cases with isothermal porous medium and conjugate
heat transfer were analysed. The strengthening of thermal fluctuations in
porous medium, previously attributed to the span wise rollers, have been found
to be caused also by a strong averaged temperature gradient in the porous
medium.

The difference between macroscopic and microscopic scales (i.e., device and
pore scales) is typically too large to resolve the flow field between the
elements of the solid phase in engineering simulations. To reduce the
computational effort, the flow in the porous medium is often upscaled
\citep{battiato2019}. In current work, the Volume Averaging Theory (VAM) is
used, which involves a filtering procedure similar to that
adopted in \LES{}, resulting in equations for the spatially averaged equations
\citep{whitaker1999,battiato2019,quintard2016}. VAM has been successful in a
broad range of applications, in particular, turbulent flows at the PFI have
been studied with the upscaled approach, where, both DNS \citep{breugem2005,breugem2006}
and LES \citep{sadowski2023,karra2024} was used to model turbulence.

In VAM, the choice of filtering molecule is inherently connected to a chosen
porous medium \citep{quintard1993}, and it has been showed to strongly impact
the distribution of unclosed terms near the interface
\citep{sadowski2023,sadowski2023b}. Moreover, the filtering procedure typically
assumes that the use of a chosen kernel results in the filtered (macroscopic)
variables devoid of microscale fluctuations. Formally, this is expressed as a
set of constraints for the separation of micro, macro and filtering
lengthscales. When the constraints are satisfied, the filtered variables are
assumed to be constant over the support of the filtering molecule
\citep{gray1975,davit2017}, which allows for neglecting a set of terms
resulting from the Taylor series expansion of the filtered variables under the
surface filtering integral \citep{quintard1993}. \citet{battiato2019} pointed
out that such assumption leads to errors in a scalar transport case, which can
be remedied by explicitly including these terms.

Therefore, three open questions emerge from the above literature review. First,
the influence of the Prandtl number on the heat transfer at PFI has not been
systematically studied --- each work
\citep{kuwata2020,nishiyama2020a,chandesris2013} only considered one value of
$\Pr$, preventing evaluation of the influence of the temperature diffusivity
on the upscaled flow description. Second, the importance of the aforementioned
neglected terms is not assessed in turbulent heat transfer at the porous-fluid
interface. Lastly, although \citet{chandesris2013} have commented on the kernel
requirements for the filtering of the heat transfer equation, the influence of
the filtering molecule on the distributions of the unclosed quantities has not
been thoroughly mapped.

We address these questions with an analysis of wall-resolved LES of turbulent
flow with heat transfer, in a channel half-filled with cubes at porosity 0.875,
at $\Re_\mathrm{bulk} = 5485$ and $\Pr = 0.71$, $0.1$, $0.05$. The geometry and
the Reynolds number matches the DNS of \citet{breugem2005}, used as primary
reference for validation of velocity (results of the work of
\citeauthor{breugem2005} are termed hereafter BB2005). The Prandtl
numbers span the range of gaseous flow in packed beds ($\Pr = 0.71$) down to
values corresponding to heat transfer in liquid metals ($\Pr = 0.05$). We
simulate two heat-flux configurations. In the first, heat is injected only on
the cubes with top and bottom walls adiabatic, as in \citet{chandesris2013}.
The results of their work will be named C2013 in the remainder of the
text. In the second, the heat flux is also prescribed on the top wall. The two
configurations test the sensitivity of the balance terms to the boundary
conditions. For each $\Pr$ and boundary conditions configuration, we derive and
evaluate the double-averaged temperature equation \emph{via} explicit filtering
procedure using three different kernels, and we explicitly retain all terms
originating from Taylor series expansion up to a second order.

The remainder of the paper is organized as follows. \Cref{sec:Geometryflow}
describes the geometry and flow conditions. \Cref{sec:MathematicalModel}
presents the governing equations and the volume-averaging framework, while
\Cref{sec:Numericalmethod} describes the numerical method. \Cref{sec:Results} reports
the results, and lastly in \Cref{sec:Conclusions} we offer discussion and
summary of the conclusions.

\section{Geometry \& flow description}
\label{sec:Geometryflow}
\begin{figure}[t!]
	\begin{center}
		\includegraphics[width=8cm]{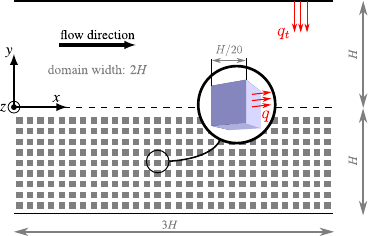}
	\end{center}
	\caption{The schematic of the studied geometry. The gray squares indicate
		positions of the cubes. The top wall and the cubes are heated with a heat
		flux of $q$ and $q_t$, respectively.}\label{fig:geometry}
\end{figure}

\begin{figure}[t!]
	\begin{center}
		\includegraphics[width=8cm]{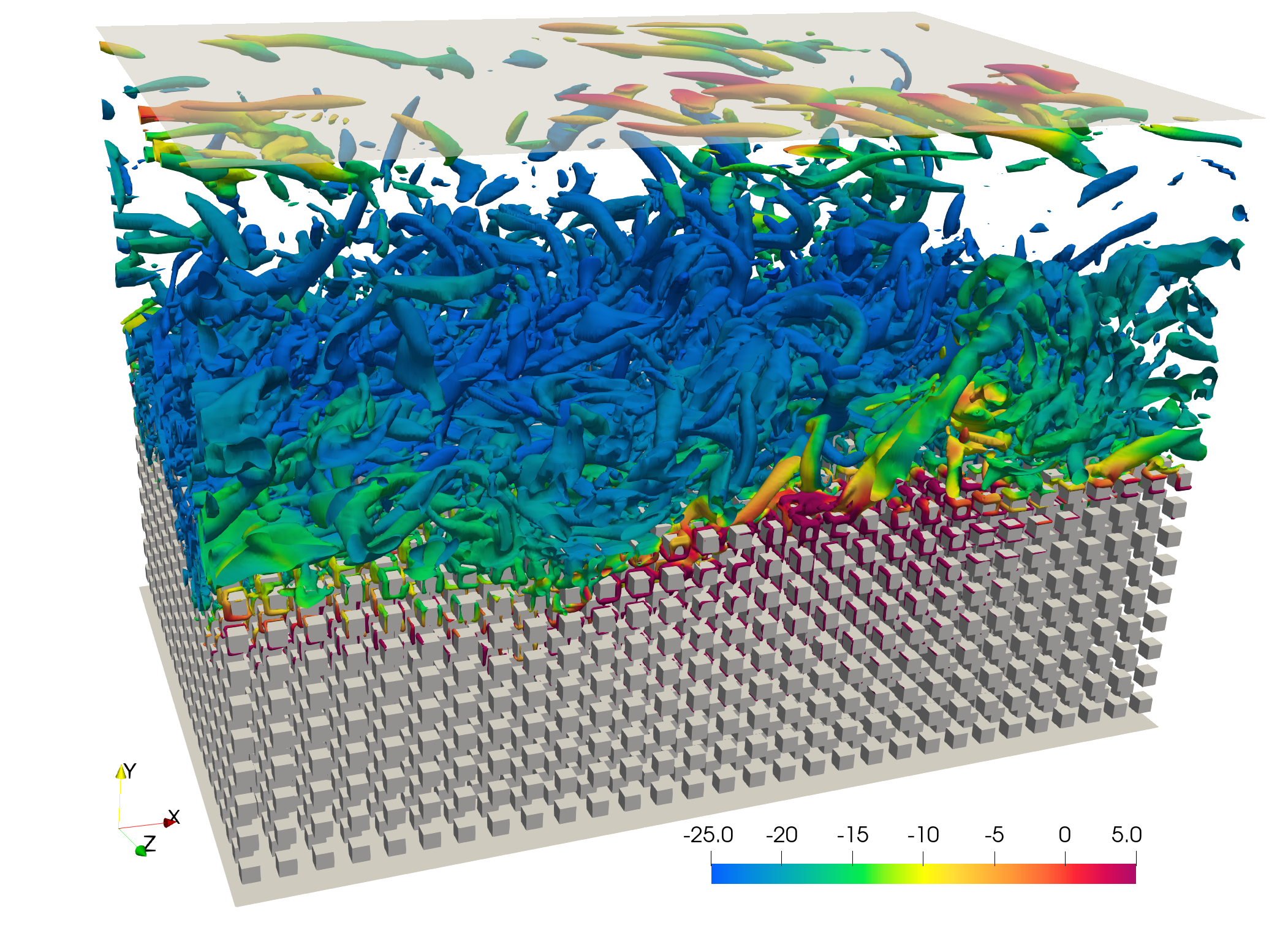}
	\end{center}
	\caption{Overview of the geometry and the turbulent flow above the porous
		wall for the case with the heated top wall. The vortical structures are
		visualized with iso-surfaces of the $Q$ criterion
		($Q\approx0.6U_\mathrm{bulk}^2/H$) and colored by
		$\theta^{p+}$ ($\Pr=0.71$).}\label{fig:overview}
\end{figure}

The geometry of the computational domain is shown in \cref{fig:geometry}. The
axes of the coordinate system, $x$, $y$ and $z$, represent the streamwise,
wall-normal, and spanwise direction, respectively. The size of the domain is
identical as in the studies of \citet{breugem2005} and \citet{chandesris2013},
resulting in a channel with dimensions $(3H \times 2H \times 2H)$ where, $H$ is
the height of the channel and the depth of the porous region. The domain is
periodic in both streamwise and spanwise directions.

At the bottom of the channel 5400 cubes with the size of $d_p = H/20$ are
arranged in 9 layers, and are spaced every $H/10$ in each direction (see
\cref{fig:overview}). The centres of the cubes in the lowest layer are located
$3H/40$ from the bottom wall. The nominal porosity (defined using the
representative periodically repeating volume containing one cube) of the
resultant porous matrix is $\phi_c =
	0.875$. The nominal permeability can be characterized by a Darcy number
$\mathrm{Da}_c = K_c/H^2 = \num{3.4e-4}$, where $K_c$ is the permeability
computed for a homogenous porous medium consisting of such array of cubes
\citep{breugem2004}.

The flow conditions in the clear channel are defined by the Reynolds number
$\mathrm{Re}_\mathrm{bulk} = U_\mathrm{bulk}H/\nu$, where $U_\mathrm{bulk}$ is
the average velocity for $y>0$, or $\mathrm{Re}_\tau^p = u_\tau^p H / \nu =
	684$, where $u_\tau^p$ is the friction velocity over the permeable wall of the
channel. It can be computed using the global force balance in the clear channel
\citep{kuwata2016},
\begin{equation}
	u_\tau^p = \sqrt{-{(u_\tau^t)}^2 + \Pi H},
	\label{eq:u_tau_p}
\end{equation}
where $u_\tau^t=\sqrt{\tau_w(y=H)}$ is the friction velocity on the top wall
and $\Pi$ is the mean pressure gradient driving the flow. Hereafter, the
superscripts ${(\bm{\cdot})^{p+}}$ and ${(\bm{\cdot})^{t+}}$ represents values
normalized using $u_\tau^p$ and $u_\tau^t$, respectively.
Such Reynolds number results in fully developed turbulent flow in the channel.

For the ease of reproducibility of the
simulated flow conditions, a total Reynolds number can also be defined as
$\mathrm{Re}_\mathrm{tot} = 2U_\mathrm{tot}H/\nu$,
where $U_\mathrm{tot}$ is the average velocity in the whole domain. The
non-dimensional numbers for all conducted simulations along with the reference
DNS are gathered in \cref{tab:simulations}. The considered flow conditions
correspond to turbulent flow in the clear part of the channel and laminar flow
between the cubes.

The work of \citet{chandesris2013} explored different configurations of
boundary conditions for heat transfer in the present geometry and concluded
that, although certain closure coefficients, like tortuosity coefficient, or
turbulent heat transfer inside the porous medium are not strongly impacted by
the choice of boundary conditions, the distributions of temperature and its
variance are strongly affected. For this reason, in current work we have decided
to consider two different configurations of boundary conditions for heat transfer.
First one corresponds to the study of \citet{chandesris2013}, where
a uniform unit heat flux $q$ is prescribed on the surface of the cubes and the
top and bottom wall of the channel are treated as an adiabatic boundary. In
the second one, the heat flux $q_t$ is also prescribed on the top wall of the
channel, so that the total heat flux through the top wall is equal to the total
heat flux through the cubes.

The previous study of heat transfer in the considered geometry
only considered a Prandtl number of $\mathrm{Pr} = 0.1$
due to a large time required for establishing a fully developed solution and
subsequent averaging of temperature fields and its moments. Following
\citet{chandesris2013}, the ratio between the time scales of thermal diffusion
in the porous region and the turbulent diffusion of momentum in the clear part
of the channel can be estimated as $\mathrm{Re}^p_\tau \mathrm{Pr}$, which for
$\mathrm{Pr}=0.1$ is around $68$.
In the current study, for each set of temperature boundary conditions, we study
heat transfer for three different Prandtl numbers $\mathrm{Pr}=0.71, 0.1,
	0.05$. The highest value of $\mathrm{Pr}$ results in the timescale ratio around
485.

To non-dimensionalise the fluctuations of temperature fields it is useful to
define the friction temperature in a manner similar to the friction velocity.
From the total energy balance in the clear channel, we can evaluate the heat
flux at the porous-fluid interface as
\begin{equation}
	q_p = -q_t + \gamma H U_\mathrm{bulk}
	\label{eq:q_p}
\end{equation}
and the friction temperature is defined as $\theta_\tau^p = q_p / u_\tau^p$
\citep{nishiyama2020a}, where $\gamma$ represents the coefficient of the linear
increase of temperature in the channel.

\begin{table*}
	\caption{The macroscopic variables characterizing each conducted simulation
		and the reference DNSs by \citet{breugem2005} labelled as BB2005, and by
		\citet{chandesris2013} labbeld as C2013. The Reynolds
		numbers are defined as
		$\mathrm{Re}_\mathrm{bulk} = U_\mathrm{bulk}H/\nu$,
		$\mathrm{Re}_\mathrm{tot} = 2U_\mathrm{tot}H/\nu$,
		$\mathrm{Re}_\tau^p = u_\tau^p H/\nu$,
		where $U_\mathrm{tot}$ and $U_\mathrm{bulk}$ are
		average velocities in the whole domain and in the clear channel, respectively. The,
		skin friction coefficients are defined as $C^{p,t}_f = 2 {(u^{p,t}_\tau/U_\mathrm{bulk})}^2$.
	}\label{tab:simulations}
	\begin{center}
		\begin{tabular}{@{}lllllll@{}}
			\toprule
			Simulation & $\mathrm{Re}_\mathrm{tot}$ & $\mathrm{Re}_\mathrm{bulk}$ & $\mathrm{Re}_{\tau}^p$ & $\mathrm{Re}_K$ & $C_f^t \times 10^{-3}$ & $C_f^p\times 10^{-3}$ \\ \midrule
			current    & 6360                       & 5485                        & 684                    & 12.4            & 10.280                 & 31.173                \\
			BB2005     & -                          & 5500                        & 669                    & 12.4            & 10.264                 & 29.6                  \\
			C2013      & -                          & 5423                        & 664                    & -               & -                      & -                     \\
			\bottomrule
		\end{tabular}
	\end{center}
\end{table*}

\section{Mathematical model}
\label{sec:MathematicalModel}
We are considering a fully-developed turbulent flow in a periodic channel,
under low-\Ma{}, isochoric conditions.
To take into account the periodicity of the computed geometry the kinematic
pressure $p$ (i.e., the static pressure divided by the density) is decomposed
into the periodic component $\per{p}$ and a linear change in pressure in the
streamwise direction $x_1$, prescribed as a pressure drop over the domain
length $L$:
\begin{equation}
	p = \per{p} + \dfrac{\Delta p}{L}x_1 = \per{p} - \Pi x_1,
\end{equation}
resutling in the following conservation law for momentum
\begin{equation}
	\pdv{u_i}{t} + u_j \pdv{u_i}{x_j} = -\pdv{\per{p}}{x_i} + \nu\pdv{u_i}{x_j,x_j} + \Pi\delta_{i1},
	\label{eq:momentum}
\end{equation}
where $\vec{u}$ is the velocity and $\nu$ is the fluids kinematic viscosity. The
conservation of mass is governed by the continuity equation
\begin{equation}
	\pdv{u_i}{x_i} = 0.
	\label{eq:continuity}
\end{equation}

The temperature field $\theta$ can be decomposed similarly, into the periodic
component $\per{\theta}$ and the linear rise of $\theta$ due to the heat flux $q$ through a surface $A_q$
\citep{chandesris2013},
\begin{equation}
	\theta = \per{\theta} + \dfrac{q A_q }{V U_\mathrm{bulk}}x_s
	=  \per{\theta} + \gamma x_1,
	\label{eq:scalar-per-decomp-expl}
\end{equation}
where $U_\mathrm{bulk}$ is the average velocity in the streamwise direction and
$V$ corresponds to total volume of fluid in the domain. When the temperature is
approximated as a passive scalar, the evolution of the periodic component of $\theta$ is
than governed by the equation
\begin{equation}
	\pdv{\per{\theta}}{t}
	+ u_i \pdv{\per{\theta}}{x_i}
	= \alpha \pdv{\per{\theta}}{x_i,x_i}
	- u_1 \gamma
	\label{eq:scalar-eq}
\end{equation}
where $\alpha$ is the thermal diffusivity coefficient which forms the Prandtl
number $\Pr = \nu / \alpha$.

When Large Eddy Simulation (LES) is conducted, the filtered equations are solved instead of
\cref{eq:momentum,eq:continuity,eq:scalar-eq} \citep{sagaut2006}. The filtering operation,
denoted here with $\fil{\bm{\cdot}}$, results in filtered continuity equation
$\pdv{\fil{u}_i}/{x_i} = 0$
and momentum equations
\begin{equation}
	\pdv{\fil{u}_i}{t} + \fil{u}_i \pdv{\fil{u}_i}{x_j}  =
	-\pdv{\fil{\per{p}}}{x_i}
	+ \nu\pdv{\fil{u}_i}{x_j,x_j}
	+ \pdv{}{x_j}\ab(2\nu_\sgs\fil{\tnsr*{S}}_{ij}) + \Pi \delta_{i1}
	\label{eq:les-mom}
\end{equation}
where eddy viscosity model has been used for the closure of the sub-filter-stresses and
$\fil{\tnsr*{S}}_{ij} = (\partial \fil{u}_i /\partial x_j + \partial\fil{u}_j /\partial x_i)/2$.
Similarly, the heat transfer equation after filtering and closure becomes
\begin{equation}
	\pdv{\fil{\per{\theta}}}{t}
	+ \fil{u}_i \pdv{\fil{\per{\theta}}}{x_i}
	= \alpha \pdv{\fil{\per{\theta}}}{x_i,x_i}
	+ \pdv{}{x_j}\left( \frac{\nu_\sgs}{\Pr_t} \pdv{\fil{\per{\theta}}}{x_j} \right)
	- \fil{u}_1 \gamma.
	\label{eq:les-scalar-eq-per}
\end{equation}
The turbulent Prandtl number $\Pr_t$ is often taken as $0.85$ for
gases \citep{pope2000}. We adopt this value for each of the considered Prandtl
number --- for turbulent heat transfer at low $\mathrm{Pr}$ the influence of sub-grid
model should be weak due to the smaller separation of scales. For the sake of
clarity, in the remainder of this work we will drop the $\fil{\bm{\cdot}}$ and
$\per{\bm{\cdot}}$ accents.

\subsection{Volume Averaging Theory}
A detailed overview of various upscaling approaches (i.e., deriving governing
equations for macroscale process from the equations guiding the microscale) is
available in the reviews by \citet{battiato2019} and \citet{davit2013}. In the
current work we employ the Volume Averaging Theory (VAT) which has been
established by \citet{whitaker1999} and \citet{quintard1993}.
In this framework, to obtain a macroscopic description of the flow properties a
volume averaging (of spatial filtering) operation has to be performed on the
gathered microscopic data. The volume average of a variable defined in the
fluid phase can be formulated by introducing an appropriate weighting function
(or filtering kernel) and representing the average as a filtering operation
similar to LES filtering:
\begin{equation}
	\va_{\psi}(\vec{x}) =
	\int\limits_{\mathbb{R}^3}G(\vec{x}-\vec{\xi})I(\vec{\xi}) \psi(\vec{\xi}) \odif{\vec{\xi}},
	\label{eq:vam-sup-avg-filtering}
\end{equation}
where $I$ is a phase indicator function \citep{gray1975}
\begin{equation}
	I(\vec{x}) = \left\{
	\begin{array}{@{}l@{\thinspace}l}
		1 & : \text{for }\vec{x} \text{ in fluid;} \\
		0 & : \text{otherwise.}                    \\
	\end{array}
	\right.
	\label{eq:VAM:gamma}
\end{equation}

\begin{figure}
	\begin{center}
		\includegraphics[width=8cm]{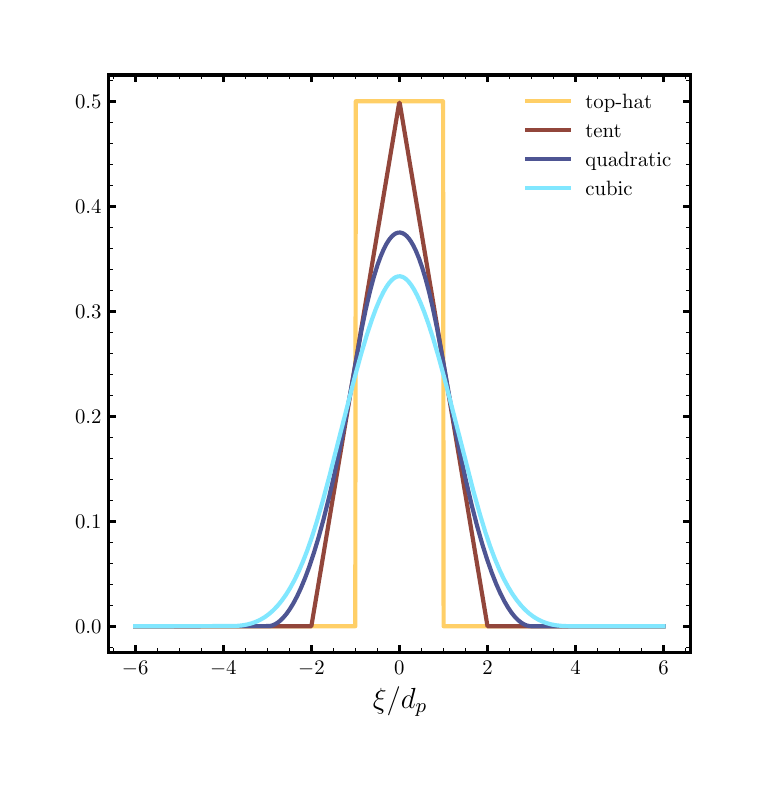}
	\end{center}
	\caption{The filtering kernels used in the study, tent:
		\cref{eq:tent-kernel}, quadratic: \cref{eq:quadratic-kernel}, cubic:
		\cref{eq:cubic-kernel}, along with the top-hat kernel used to construct
		them.}\label{fig:kernels}
\end{figure}

Following the discussion in \citep{battiato2019}, the choice of the filtering
kernel is connected to the morphology of the porous medium. For
spatially periodic, ordered porous media a kernel corresponding to so-called
\emph{cellular} average should be used \citep{quintard1993}, to filter out
linear component of variations of microscopic fields.
Such a kernel is formed by taking a convolution of a top-hat kernel with
itself, resulting in a \say{tent} function (depicted in \cref{fig:kernels}),
which can be written as
\begin{equation}
	G_1(\xi)=
	\left\{
	\begin{array}{@{}l@{\thinspace}l}
		\dfrac{\ell - |\xi|}{\ell^2} & : \text{for }|\xi| \leq \ell \\
		0                            & : \text{otherwise.}          \\
	\end{array}\right.
	\label{eq:tent-kernel}
\end{equation}
where $\ell$ is the kernel half-size and corresponds to the width of the
original top-hat kernel.

This process can be repeated to filter out higher
order terms if needed, as by \citet{dhueppe2011,chandesris2013}, who pointed out that a triple
convolution of a top hat function resulting in a second-order (or
\say{quadratic}) filtering kernel is required to remove the fluctuations from the gradients of
the filtered quantities. This filtering molecule can be formulated as:
\begin{equation}
	G_2(\xi)=
	\left\{
	\begin{array}{@{}l@{\thinspace}l}
		\dfrac{1}{\ell}\ab(\dfrac{3}{4} - \dfrac{\xi^2}{\ell^2})                        & : \text{for } -\frac\ell2 \leq \xi \leq \frac\ell2   \\
		\dfrac{1}{2\ell}\ab(\dfrac{\xi^2}{\ell^2} -\dfrac{3|\xi|}{\ell} + \dfrac{9}{4}) & : \text{for }  \frac\ell2 < |\xi| \leq \frac{3\ell}2 \\
		0                                                                               & : \text{otherwise.}                                  \\
	\end{array}\right.
	\label{eq:quadratic-kernel}
\end{equation}
Depending on the evaluated quantities,
further regularization of the macroscopic variables might be required. A
\say{cubic} kernel can be constructed by performing a quadruple convolution of
the top-hat filtering molecule resulting in
\begin{equation}
	G_3(\xi)=
	\left\{
	\begin{array}{@{}l@{\thinspace}l}
		\dfrac1{\ell}\ab(\dfrac{2}{3} - \dfrac{\xi^2}{\ell^2} + \dfrac{{|\xi|}^{3}}{2\ell^3})                     & : -\ell \leq \xi \leq \ell \\
		\dfrac4{3\ell}\ab(1 - \dfrac{3|\xi|}{2\ell}  + \dfrac{3\xi^{2}}{4\ell^2}  - \dfrac{{|\xi|}^{3}}{8\ell^3}) & : \ell < |\xi| \leq 2\ell  \\
		0                                                                                                         & : \text{otherwise.}        \\
	\end{array}\right.
	\label{eq:cubic-kernel}
\end{equation}

An extensive discussion of kernel properties can be found in \citep{davit2017}.
The kernel has to preserve constants, i.e., it has to be normalized such that
its integral is equal to $1$, and be symmetric, i.e., \(G(\vec{x}) =
G(-\vec{x})\). Moreover, \(G\in C^\infty\) is required, which
can be achieved for kernels similar to \cref{eq:tent-kernel}
by taking a convolution between the chosen kernel and $C^\infty$
compactly-supported mollifying function with much smaller characteristic size.
This results in $C^\infty$ kernel function, resembling the original one
\citep{battiato2019}. That being said, this additional \say{smoothing}
operation is rarelly performed in
practice when macroscopic fields are evaluated from microscopic simulation
data \citep[an example of such proccess avialble in the work of][]{battiato2019}.
On the other hand, \cref{eq:tent-kernel,eq:quadratic-kernel} or similar
expressions which are only piecewise infinitely differentiable have been often used
\citep{sadowski2023b,breugem2005,chandesris2013}.
Additionally, $G$ is expected to have a compact support
--- an assumption which could perhaps be restated as requirement for a fast
decay of $G(\xi)$ with increasing $\xi$, as \citet{sadowski2023b} successfully
used Gaussian kernel when considering the laminar flow at the PFI.

Using \cref{eq:vam-sup-avg-filtering}, the porosity can be computed as $\phi = \va_{1}$
and, using porosity, and intrinsic average can be defined
\begin{equation}
	\phi\va{\psi} = \va_{\psi},
	\label{eq:vam-int-avg}
\end{equation}
together with the corresponding decomposition of a variable defined in the fluid
into the \say{mean} and \say{fluctuating} parts
\begin{equation}
	\psi = \va{\psi} + \va*{\psi}.
	\label{eq:decomp}
\end{equation}

In VAM, for stationary porous medium the filtering commutes with time
derivatives, however the spatial derivatives are no longer commuting with
filtering operation due to the solid bodies inside the support of the kernel.
To derive the upscaled equations, the \emph{spatial} or \emph{general}
averaging theorem should be used \citep{whitaker1999}. The theorem was first
derived assuming averaging over a representative volume \citep{howes1985} and
was later generalized to allow for the use of arbitrary kernels
\citep[derivation of theorems avialable
	in][]{gray1989,quintard1994b}. The spatial theorem connects average
of a gradient with a gradient of an average
\begin{equation}
	\va_{\pdv{\psi}{x_i}} = \pdv{\va_{\psi}}{x_i} +
	\int\limits_A G n_i \psi \odif{A}
	\label{eq:VAM:sp-theorem}
\end{equation}
where $A$ is the wetted surface of the solid and \(\vec{n}\) is the surface
normal of $A$ oriented into the solid.\footnote{%
	This choice is arbitrary: if outwards normal is chosen, a change of signs in
	both theorems is required.
} The surface integral over $A$ is often referred to as a surface filter.
By substituting \(\psi=1\) into \cref{eq:VAM:sp-theorem},
we can derive also another useful relation:
\begin{equation}
	\pdv{\phi}{x_i} = - \int\limits_A G n_i \odif{A},
	\label{eq:VAM:avg-sp-porosity}
\end{equation}
which can be further generalized \citep{quintard1994b}
to a family of \emph{geometric} averaging theorems
\begin{equation}
	\pdv{}{x_i}{\va_{\xi_\alpha\xi_\beta\xi_\gamma\dots}} =
	- \int\limits_A  G n_i\xi_\alpha\xi_\beta\xi_\gamma\dots \odif{A} .
	\label{eq:VAM:geom-theorem}
\end{equation}
These theorems are useful as the upscaling procedure often involves surface
filtering of spatially filtered variables, with the products of $\vec{\xi}$
appearing as a result of Taylor expansion. If \(\va{\psi}\) can be expanded
into a Taylor series around \(\vec{x}\), over the support of \(G\)
\begin{equation}
	\va{\psi}(\vec{x} + \vec{\xi}) =
	\va{\psi}(\vec{x})
	+ \xi_i\pdv{\va{\psi}}{x_i}\big\vert_{\vec{x}}
	+ \frac{1}{2}\xi_i \xi_j \pdv{\va{\psi}}{x_i x_j}\big\vert_{\vec{x}}
	+ \dots,
	\label{eq:VAM:Taylor-exp}
\end{equation}
using \cref{eq:VAM:avg-sp-porosity,eq:VAM:geom-theorem} these surface integrals
can be rewritten as:
\begin{equation}
	\begin{split}
		-\int\limits_A G n_k \va{\psi}\odif{A}
		= & \va{\psi}\pdv{\phi}{x_k}
		+  \pdv{\va_{\xi_i}}{x_k}\pdv{\va_{\psi}}{x_i}                           \\
		  & +  \frac{1}{2} \pdv{\va_{\xi_i \xi_j }}{x_k}\pdv{\va{\psi}}{x_i,x_j}
		+ \dots,
	\end{split}
	\label{eq:VAM:simplified-surf-avg}
\end{equation}
since the terms evaluated at \(\vec{x}\) are constant
while evaluating the convolution integral.

For ordered spatially periodic
porous media, \citet{quintard1994b,quintard1994d} proved that
\begin{equation}
	\langle\underbrace{\xi_i\xi_j\xi_k\dots}_{n \;\text{times}}\rangle_s = \left\{
	\begin{array}{@{}l@{\thinspace}l}
		0               & : \text{when }n\text{ is odd}   \\
		\text{constant} & : \text{when }n\text{ is even,} \\
	\end{array}
	\right.
	\label{eq:VAM:geometric-ordered}
\end{equation}
as long as \(G\) is defined as cellular average \citep{quintard1993}, which leads
to the simplified version of \cref{eq:VAM:simplified-surf-avg}
\begin{equation}
	-\int\limits_A n_k \va{\psi}\odif{A}
	= \va{\psi}\pdv{\phi}{x_k}.
	\label{eq:VAM-approx}
\end{equation}

\subsection{Double-averaged equations}

To enable analysis and modelling of turbulence near porous regions, the
governing equations need to be first averaged in time, which is indicated with the overbar ($\overline{\;\bm{\cdot}\;}$), and subsequently averaged
in space using the VAT framework. This leads to so-called double-averaged
equations \citep{lage2002,nikora2007,sadowski2023}. Here, we briefly introduce
the equations governing the flow \citep[a detailed derivation together with the
	discussion of the closure is available in the works
	by][]{breugem2005,breugem2006,whitaker1996}, which is followed by the full
derivation of the scalar equations.

Applying averages and averaging theorems to \cref{eq:continuity}
results in a following continuity equation
\begin{equation}
	\pdv{\phi\da{u_i}}{x_i} = 0,
	\label{eq:va-cont}
\end{equation}
where we also implicitly assume that the porous medium is stationary. We can
define the material derivative as
$\mdv{\psi}/{t} = \pdv{\psi}/{t} + \da{u_i}\pdv{\psi}/{x_i}$
and using this definition, the double-averaged momentum equation reads
\begin{equation}
	\begin{split}
		\phi\mdv{\da{u_i}}{t}
		=
		- \phi\pdv{\da{p}}{x_i}
		+ \phi\nu\pdv{\da{u}}{x_j,x_j}
		+ \nu \pdv{\phi}{x_j}\pdv{\da{u_i}}{x_j} \quad \\
		+ \nu\da{u_i}\pdv{\phi}{x_j,x_j}
		+ \pdv{\phi\tnsr*{T}_{ij}}{x_j}
		+ F_i + E_i^{u} + \vec{\mathcal{H}}_i^u
		+ \phi\Pi\delta_{i1}.
	\end{split}
	\label{eq:va-mom}
\end{equation}
where $\phi$ is the porosity, $\tnsr{T}$ corresponds to filtering induced stress
\begin{equation}
	\tnsr{T} = \da{\bm{u}}\da{\bm{u}} - \da{\bm{u}\bm{u}} =
	\underbrace{\da{\bm{u}}\da{\bm{u}} - \va{\overline{\bm{u}}\,\overline{\bm{u}}}}_{\text{dispersion}} - \underbrace{\da{\bm{u}'\bm{u}'}}_\text{turbulent}
	\label{eq:non-lin-decom}
\end{equation}
containing both spatial average of Reynolds stresses $\da{\bm{u}'\bm{u}'}$ and
the momentum dispersion tensor, a contribution of sub-filter scales of
time-averaged velocity. The latter is often neglected inside the porous medium
due to its low contribution to momentum balance \citep{breugem2006}. The drag
force induced by the presence of the solid elements is given by
\begin{equation}
	F_i =
	\int\limits_{A}G
	\left(
	\va*{\overline{p}}\delta_{ij} + \nu\pdv{\va*{\overline{u_i}}}{x_j}
	\right)n_j
	\odif{A},
	\label{eq:uncl-drag}
\end{equation}
and we collect the terms related to explicit LES influence\footnote{%
	Although, by definition $\nu_\mathrm{sgs}=0$ at the solid viscous wall, we
	retain the surface integral of $\overline{\nu_\mathrm{sgs}\tnsr{S}}$, since
	it's value is actually implementation specific. In OpenFOAM, the value at the
	wall is specified by a wall function and is small in magnitude but not $0$.}
under $\vec{E}^u$
\begin{equation}
	E_i^{u} = \int\limits_{A}G
	\overline{\nu_\sgs\tnsr*{S}_{ij} }
	n_j
	\odif{A} + \pdv{ \phi\da{\nu_\sgs\tnsr*{S}_{ij} }}{x_j}.
	\label{eq:mom-LES}
\end{equation}
In a well resolved LES, this term should not contribute significantly to the
overall double-averaged momentum balance, as already tested by
\citet{sadowski2023}. Lastly, $\vec{\mathcal{H}}^u$ results from applying the Taylor
expansion of double-averaged variables under the surface integrals and
subsequent use of
\cref{eq:VAM:geom-theorem,eq:VAM:simplified-surf-avg}:
\begin{equation}
	\vec{\mathcal{H}}^u_i = \pdv{\va_{\xi_\alpha}}{x_i}\pdv{\da{p}}{x_\alpha} + \dots
	- \nu\pdv{\va_{\xi_\alpha}}{x_j}\pdv{\da{u_i}}{x_\alpha,x_j} - \dots
	\label{eq:HOT-momentum}
\end{equation}
In most derivations of double-averaged momentum equation, these terms are
neglected, i.e., \cref{eq:VAM-approx} is used directly, based on the assumption
of the separation of scales (for unordered porous medium) or by selecting
appropriate kernel for a chosen ordered porous medium \citep{quintard1993}. On
the interface, however, the separation of scales is more difficult to define
and especially \cref{eq:VAM:geometric-ordered} no longer holds, as the porous
medium inside the kernel's support is no longer periodic.

In a statistically one-dimensional flow as in the present case, only
streamwise component of $\vec{\mathcal{H}}^u$ is significant and only derivatives with
respect to wall-normal coordinate $y$ are not equal to 0, resulting in:
\begin{equation}
	\mathcal{H}^u_x =- \nu\pdv{\va_{\xi_2}}{y}\pdv{\da{u}}{y,y} + \dots
\end{equation}
These terms were a focus of work of \citet{goyeau1997}, who showed that for porous
structures with a smooth variation of porosity these terms will not
significantly contribute to the momentum balance. Indeed, the gradient of
$\va_{\vec{\xi}}$ can be estimated as $\mathcal{O}\ab(1-\phi)$
\citep{quintard1993} which motivates neglecting the term in highly porous and
permeable regions. For turbulent flows near the interface, the contribution of
$\nu\pdv[order=2]{\da{u}}/{y}$ towards the momentum budget is much smaller
\citep{sadowski2023} than that of the drag force or volume averaged Reynolds
stresses, which also allows for the simplification. Therefore, this term can be
safely omitted in typical cases involving highly permeable walls, while it
might be significant for laminar flows or flows near moving porous matrices.

For the temperature equation the derivation is following similar steps
\citep{quintard2016,battiato2019}. Applying time averaging, spatial filtering
and \cref{eq:VAM:sp-theorem} to \cref{eq:les-scalar-eq-per} results in
\begin{equation}
	\begin{split}
		\pdv{\phi\da{\theta}}{t}
		+ \pdv{\phi \da{u_i \theta}}{x_i}
		= \alpha \pdv{\phi\da{\theta}}{x_i,x_i}
		+ \pdv{}{x_i}\da{\frac{\nu_{sgs}}{\Pr_t} \pdv{\theta}{x_i}} \\
		+
		\int\limits_A G \ab(
		\alpha\pdv{\overline{\theta}}{x_i}
		+\overline{\frac{\nu_{sgs}}{\Pr_t} \pdv{\theta}{x_i}}
		)
		n_i\odif{A}
		\\
		+ \alpha\pdv{}{x_i} \int\limits_A G \theta n_i\odif{A}
		- \phi \da{u_1} \gamma.
	\end{split}
\end{equation}
First, the terms related to the sub-grid scale viscosity can be again grouped
together
\begin{equation}
	E^\theta = \pdv{}{x_i}\da{\frac{\nu_{sgs}}{\Pr_t} \pdv{\theta}{x_i}} +
	\int\limits_A G
	\overline{\frac{\nu_{sgs}}{\Pr_t} \pdv{\theta}{x_i}}
	n_i\odif{A}
\end{equation}
to represent the influence of the sub-grid modelling.
Next, the left-hand side can be rearranged
\begin{equation}
	\begin{split}
		\pdv{\phi\da{\theta}}{t}
		+ \pdv{\phi \da{u_i \theta}}{x_i} =
		\phi\mdv{\da{\theta}}{t} - \pdv{\phi q_i}{x_i},
	\end{split}
\end{equation}
where the sub-filter heat flux $\vec{q}$ contains both the volume averaged turbulent heat flux
and the dispersion heat flux:
\begin{equation}
	\vec{q} = \da{\vec{u}}\da{\theta} - \da{\vec{u}\theta}=
	\da{\vec{u}}\da{\theta} - \va{\overline{\vec{u}}\;\overline{\theta}} - \da{\vec{u}'\theta'}
	\label{eq:sub-filter-heat-flux}
\end{equation}
Similarly to the momentum dispersion stresses, \citet{chandesris2013} have reported that
the $\da{\vec{u}}\da{\theta} - \va{\overline{\vec{u}}\;\overline{\theta}}$ contributes
negligible amounts to the flux balance and can therefore be neglected.

The diffusion term can be rearranged as
\begin{equation}
	\alpha\pdv{\phi\da{\theta}}{x_i, x_i}
	= \alpha\phi\pdv{\da{\theta}}{x_i,x_i} + 2 \alpha\pdv{\da{\theta}}{x_i}\pdv{\phi}{x_i}
	+ \alpha\da{\theta}\pdv{\phi}{x_i,x_i},
\end{equation}
where the term $\alpha\pdv{\phi}/{x_i}\pdv{\da{\theta}}/{x_i}$ is called
the Brinkman term, and as it is proportional to porosity gradients,
it is only active near the interface.

The surface integral of $\pdv{\overline{\theta}}/{x_i}$ over the cubes can be,
in our case, represented exactly, therefore it will remain in the equation as
is, forming the \say{volumetric heating} term
\begin{equation}
	Q =
	\int\limits_A G
	\frac\nu\Pr\pdv{\overline{\theta}}{x_i}
	n_i\odif{A}.
	\label{eq:heat-source}
\end{equation}
The second surface integral term, on the other
hand requires further work --- the time-averaged temperature has to be split into
the spatial mean and fluctuating part, where the former is converted into spatial-filtered
quantities using \cref{eq:VAM:simplified-surf-avg}:
\begin{equation}
	\begin{split}
		 & \pdv{}{x_i} \int\limits_A G \theta n_i\odif{A}
		=
		\pdv{}{x_i} \int\limits_A G \va*{\overline{\theta}} n_i\odif{A} \\
		 & -\pdv{}{x_i} \ab[
			\da{\theta}\pdv{\phi}{x_i}
			+ \pdv{\va_{\xi_\alpha}}{x_i}\pdv{\da{\theta}}{x_\alpha}
			+ \frac12\pdv{\va_{\xi_\alpha \xi_\beta}}{x_i}\pdv{\da{\theta}}{x_\alpha, x_\beta}
			+ \dots
		].
	\end{split}
\end{equation}
The term
\begin{equation}
	\tau = \pdv{}{x_i} \int\limits_A G \alpha\va*{\overline{\theta}} n_i\odif{A}
	\label{eq:tort}
\end{equation}
is called tortuosity, and it represents the change in thermal diffusivity due
to the solid elements inside the filtering kernel
\citep{gray1975,quintard2016}. Analogously to the momentum equation, the terms
arising from the Taylor expansion can be grouped together
\begin{equation}
	\mathcal{H}^\theta = - \pdv{}{x_i}\ab(
	\alpha\pdv{\va_{\xi_\alpha}}{x_i}\pdv{\da{\theta}}{x_\alpha}
	+ \alpha\pdv{\va_{\xi_\alpha\xi_\beta}}{x_i}\pdv{\da{\theta}}{x_\alpha,x_\beta}
	+ \dots
	).
	\label{eq:high-order-theta}
\end{equation}
Their significance will be analysed in \cref{sec:hot}. Finally,
all of the above can be assembled together leading to the
double-averaged temperature equation:
\begin{equation}
	\begin{split}
		\phi\mdv{\da{\theta}}{t}
		= & \phi\frac\nu\Pr \pdv{\da{\theta}}{x_i,x_i}
		+ \frac\nu\Pr \pdv{\phi}{x_i}\pdv{\da{\theta}}{x_i} \\
		  & + \pdv{\phi q_i}{x_i}
		+ Q
		+ \tau
		- \phi\da{u_1} \gamma
		+ E^\theta + \mathcal{H}^\theta.
	\end{split}
\end{equation}

\section{Numerical method}
\label{sec:Numericalmethod}

The flow is driven by a pressure gradient, which is automatically adjusted to
ensure the correct total velocity and therefore, the total Reynolds number.
The simulation have been performed using the \textit{OpenFoam-v2312} toolbox.
For momentum equation, both spatial and
temporal derivatives have been discretized with second-order schemes. In case
of scalar equations, a \texttt{limitedLinear 0.1} scheme has been used for the
convective terms (corresponding to a bounded central difference), while all the
other terms have also been approximated with second-order schemes.

Considering only the bulk Reynolds number, the clear channel part of the domain
could be resolved without the need to model sub-grid scales. However, properly
resolving the complex geometry of the porous medium requires large
computational effort (around $60\%$ of cells are \say{placed} in the porous
region). Considering the time required for gathering statistics, a mesh
entailing 31M cells---representing a coarser resolution than the grid used for
the BB2005 DNS---was adopted.
Since, we expect
that adopted mesh resolution might not meet the DNS requirements in certain
parts of the domain, the effect of the sub-grid scales has been modelled by
means of the WALE model, as in \citep{sadowski2023} where we have also
performed a mesh sensitivity study.

\begin{table}[]
	\centering
	\caption{The maxima of near wall spacing at the top wall and the cubes.}\label{tab:avg_wallspc}
	\begin{tabular}{@{}llllll@{}}
		\toprule
		\multicolumn{3}{c}{top wall} & \multicolumn{3}{c}{cubes}                                 \\ \cmidrule(lr){1-3} \cmidrule(lr){3-6}
		$x^+$                        & $y^+$                     & $z^+$ & $x^+$ & $y^+$ & $z^+$ \\ \midrule
		2.5                          & 0.5                       & 2.5   & 7.5   & 1.4   & 6.4   \\
		\bottomrule{}
	\end{tabular}
\end{table}

\subsection{Simulation quality assessment}

LES as a numerical method is inherently influenced by a chosen grid resolution
and in the limit of DNS-like mesh resolution a well-posed LES model should
reproduce DNS result \citep{pope2000,sagaut2006}. As we neglect the influence
of sub-grid scale terms when evaluating the terms in the momentum and temperature
balances (e.g., porous induced drag), we have to ensure that the influence of the
sub-grid viscosity is sufficiently small.

For this reason, the resolution of all simulations was assessed using two LES
quality indices, the fraction of resolved turbulent kinetic energy
\citep{pope2000}
\begin{equation}
	\mathrm{IQ}_\mathrm{Pope} = \frac{\overline{u'_i u'_i}}{2k_\mathrm{sgs} + \overline{u'_i u'_i}},
	\label{eq:IQ_Pope}
\end{equation}
and the resolution index by \citet{celik2005} based on the estimated effective
Kolmogorov length scale:
\begin{equation}
	\mathrm{IQ}_\eta 
	= {\left[ 1 + \alpha_\eta{\left(
			\frac{\Delta\varepsilon_\mathrm{eff}^{1/4}}{\nu^{3/4}}
			\right)}^m\right]}^{-1},\quad \varepsilon_\mathrm{eff} = (\nu
	+\nu_\mathrm{sgs})\frac{k_\mathrm{sgs}}{(C_k\Delta)^2} .
	\label{eq:IQ_eta}
\end{equation}
In the above, the sub-grid turbulence kinetic energy is estimated as
$k_\mathrm{sgs}=\nu_\mathrm{sgs}^2/{(C_w\Delta)}^2$ where $C_w$ is the constant
of the WALE model, $C_k = 0.0376$ and $\alpha_\eta=0.05$ are empirical
constants, and $\Delta$ is the cell size (given as cube root of the volume).

For both indices, a time-averaged value above $0.8$ suggests an adequate mesh
resolution for LES. While the use of \cref{eq:IQ_eta} resulted in values
greater than $0.9$ in the whole domain, the resolved energy criterion given by
\cref{eq:IQ_Pope} was not satisfied in small regions of low velocity which is a
known feature of this quality index (for each simulation
$\overline{\mathrm{IQ}_\mathrm{Pope}} < 0.8$ was reported in less than $0.5\%$
of cells).

\begin{figure}
	\centering
	\includegraphics[width=8cm]{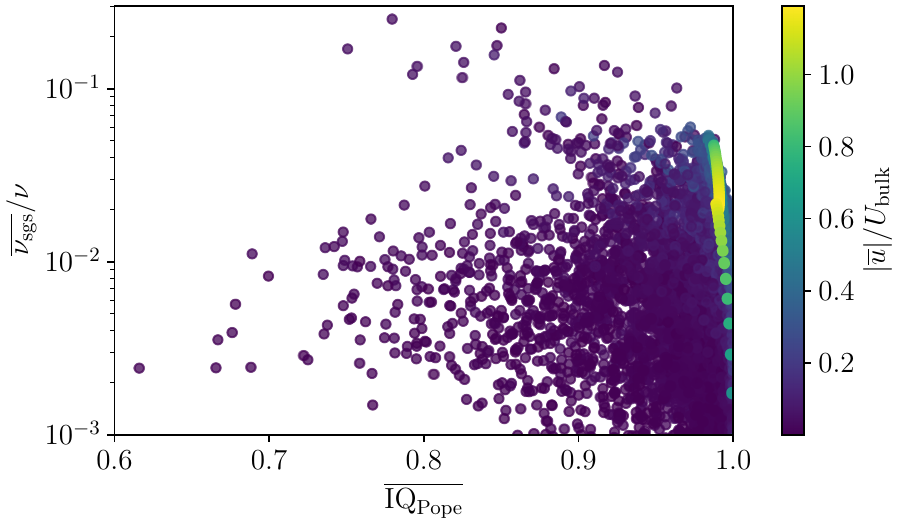}
	\caption{The scatter plot of cell $\overline{\mathrm{IQ}_\mathrm{Pope}}$
		\textit{vs.} the corresponding time-averaged contribution of sub-grid scale
		viscosity. The points are coloured by the magnitude of the time-averaged
		velocity.}
	\label{fig:quality}
\end{figure}

The scatter plot of cell $\overline{\mathrm{IQ}_\mathrm{Pope}}$ \textit{vs.}
the corresponding time-averaged contribution of sub-grid scale viscosity is
visible in \cref{fig:quality}. For most \say{under-resolved} cells, the
added viscosity is considerably smaller than the fluid's viscosity. There are,
however, a couple of cells in which
$\overline{\nu_\mathrm{sgs}}/\nu\approx0.2$. For these cells, the viscous
stresses are considerably influenced by the presence of the model, however, as
they are not numerous, we assume that they do not impact the overall simulation
fidelity.

Moreover, \cref{tab:avg_wallspc} reports the near-wall spacing at the top wall
and at the surface of the cubes, which are in line with typical recommendations
for wall-resolved LES \citep{sagaut2006, chapman1979a}.

Lastly, the size of the domain has to be large enough to allow for a proper
representation of large-scale structures present in the core flow. The chosen
length and width of the simulation domain used in this study is the same as
employed by \citet{breugem2005} and \citet{chandesris2013}. Moreover, it is
consistent with the smallest channel size found to allow for properly resolved
one-point statistics \citep{lozano-duran2014a}, however, it is typically
smaller than in other DNS studies of flows over porous substrates.
\citet{kuwata2024} used a channel with dimensions of homogeneous fluid part of
the domain given as $(6H \times H \times 3H)$. Their previous work
\citep{kuwata2016} considered dimensions of $(2\pi H \times H \times \pi H)$,
where they demonstrated the vanishing of two-point correlations
\begin{equation}
	R_{ab}(\vec{r}) =
	\frac{\overline{a'(\vec{x})b'(\vec{x}+\vec{r})}}{\overline{a'(\vec{x})b'(\vec{x})}}
	\label{eq:two-point}
\end{equation}
at $y^+\approx 20$ over approximately $1.4H$ and $0.9H$ in streamwise and
spanwise directions respectively. \citet{nishiyama2020a} demonstrated vanishing
of correlations in their much larger domain $(14H \times H \times 3.5H)$, while
noting that decreasing the streamwise length by half did not deteriorate the
second order statistics and reporting very similar shapes of scalar and
velocity correlations. Two point correlations computed for the present study
for each velocity components are plotted in \cref{eq:two-point} at three
different heights $y/H=0$, $0.1$ and $0.5$, corresponding to two planes near
the porous medium and the centre of the channel.

At $y/H=0$ and $0.1$ the correlations vanish for streamwise direction, while
the spanwise correlation of wall-normal fluctuations $R_{vv}$, flattens out
around the value of $0.1$. In the middle of the channel $R_{uu}$ does not
vanish completely, indicating that the domain is too short to capture properly
the dynamics of low-frequency structures.

Overall, the studies of the sensitivity of the gathered statistics to the chosen
domain size in the channel with smooth walls \citep{lozano-duran2014a} and in flows over
canopies similar to our configuration \citep{sathe2024}, found that smaller
domains can modify the growth of turbulent structures and
therefore, can lead to incorrectly predicted second-order statistics, while
first-order statistics typically remain mostly accurate.

The choice of the present computational domain has been dictated by the large
amount of snapshots required to converge second order statistics of the scalar
fields and availability of the reference data for this configuration, however,
two point correlations suggest that the domain is not strictly large enough to
accommodate for largest vortical structures. The shape of the computed
two-point correlations are similar to the results reported by
\citet{sathe2024}, where the computed second-order moments of velocity compared
favourably with the statistics gathered from a simulation with a larger domain,
where the velocity field was completely uncorrelated. With this in mind, the
potential inaccuracy in representation of second-order quantities in current
study has to be treated as a limitation of the chosen configuration and should
be subject to further verification in larger simulation domains.

\begin{figure}
	\centering
	\includegraphics[width=8cm]{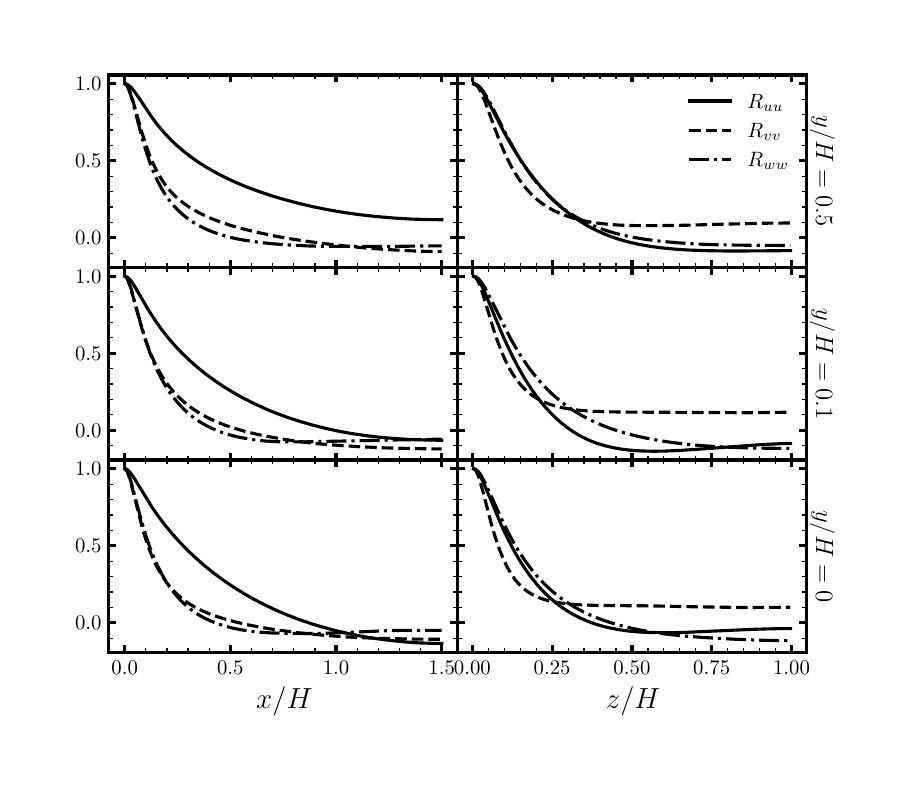}
	\caption{Two point correlations of velocity fluctuations in streamwise (left
		column) and spanwise (right column) directions, at three different heights
		($y/H = 0, 0.1, 0.5$).}
	\label{fig:two-point}
\end{figure}

\subsection{Temporal and spatial averaging}

The case was run until a velocity and temperature profiles have developed which
was tested by monitoring the convergence of mean temperatures on top and bottom
walls of the channel, as well as, in selected points inside the domain.
Snapshots of instantaneous fields were subsequently averaged over time
$2345H/U_\mathrm{bulk}$ (adiabatic top wall) and $1600H/U_\mathrm{bulk}$
(heated top wall), where sampling was conducted with the frequency of
$\sim18.2U_\mathrm{bulk}/(H)$. \citet{chandesris2013} report a longer averaging
time of $3000H/U_\mathrm{bulk}$ for their work, therefore the convergence of
gathered statistics is further checked by computing the temperature balances in the
channel, which are analysed in \cref{sec:scalar-balance}. Before the results
are filtered, they are also averaged over a periodically repeating sector of
the geometry: a box containing one column of cubes and corresponding section of
the channel.

\section{Results}
\label{sec:Results}

The overview of the flowfield from the heated top wall case is presented in
\cref{fig:overview}, where the turbulent structures are visualized over the
porous walls with the $Q$-criterion. The presence of the permeable wall results
in much stronger turbulence than the one present near the top wall
\citep{kuwata2019}. Moreover, the elongated vortical structures present neat
the top wall are not present near the permeable wall, where the turbulence is
more isotropic \citep{sadowski2023c}.

The iso-surfaces of $Q$ are coloured with the $\theta^{p+}$ at $\Pr=0.71$. The
temperature in the core flow in the centre of the channel is much colder than
the fluid inside the porous medium. The vortices near the permeable wall are
generally colder the ones in the boundary layer near the top wall, with the
comparative temperature only reached deeper inside the porous medium.

\subsection{Mean velocity and velocity fluctuations}

\begin{figure*}
	\begin{center}
		\includegraphics[width=0.95\textwidth]{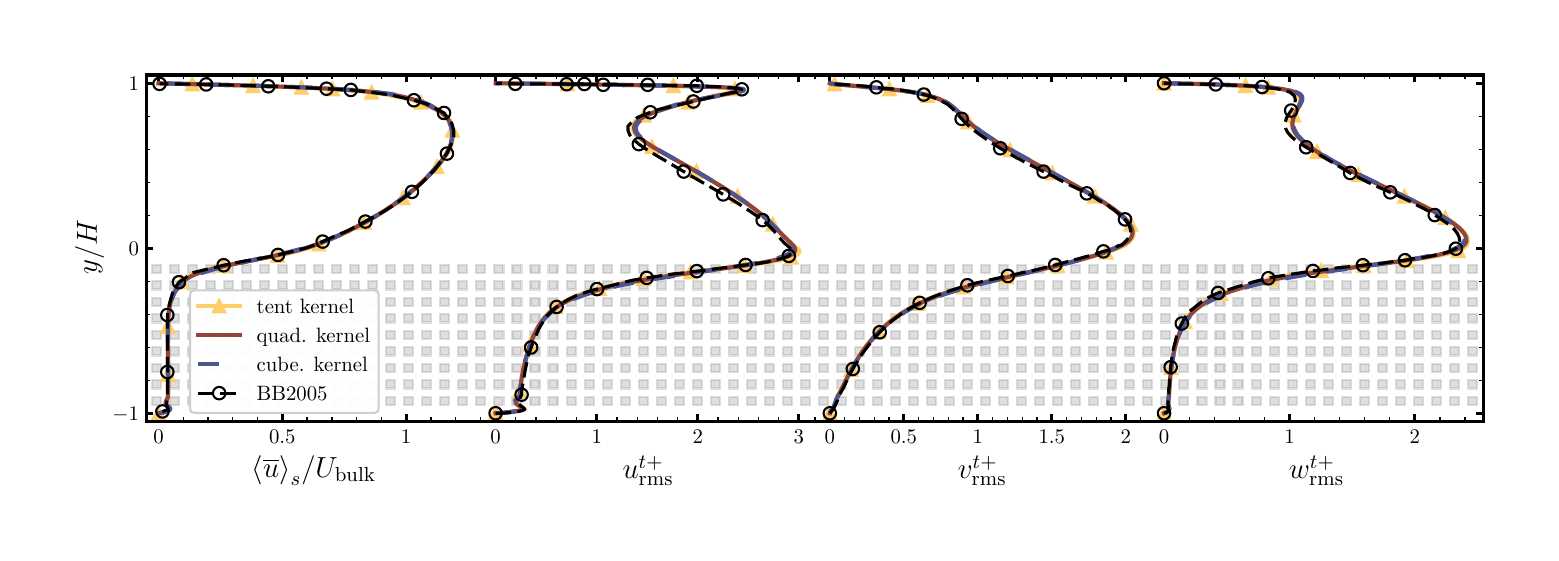}
	\end{center}
	\caption{Velocity moments in the channel (from left to right: superficial
		double-averaged streamwise velocity, root-mean-square of velocity
		fluctuations in streamwise, wall-normal and spanwise
		direction), compared to the data gathered by \citet{breugem2005}. Results with three different filtering
		kernels are presented, tent kernel given by \cref{eq:tent-kernel}, quadratic kernel \cref{eq:quadratic-kernel} and
		cubic kernel \cref{eq:cubic-kernel}.}
	\label{fig:velocity}
\end{figure*}

We can initially validate the computed results by comparing the values of the
friction coefficients
\begin{equation}
	C^{p,t}_f = 2 {(u^{p,t}_\tau/U_\mathrm{bulk})}^2
	\label{eq:friction-coeff}
\end{equation}
between our computation and the values reported by \citet{breugem2005},
which are listed in \cref{tab:simulations}. While the top wall friction
coefficient is very close to the one from the reference DNS, the friction
coefficient at the permeable wall $C^{p,t}_f$ is slightly higher.

\citet{breugem2005} evaluated $u_\tau^p$ directly from the balance of filtered
quantities as ${(u_\tau^p)}^2 = -\va_{u'w'} + \nu\pdv{\va_{u}}/{y}$. If we
follow the same procedure, we recover similar values of $C_f^p = 29.842$ and
$\mathrm{Re}_\tau^p=669.8$. In our opinion, the use of \cref{eq:u_tau_p} should
be preferred for comparison of different flow conditions, as the alternative is
inherently influenced by the filter size. The relative absolute
difference between the friction factor computed as by \citet{breugem2005} and
the one based on $u_\tau^p$ computed from total momentum balance is equal to
around $6.3\%$, $7.4\%$ and $8.3\%$ for linear, quadratic and cubic kernels
respectively.

To further validate our results we first compare the distribution of velocity and
it's second-order moments in the channel to the data gathered by
\citet{breugem2005}, who used \cref{eq:tent-kernel} for filtering. The results
are all presented in \cref{fig:velocity}.
Following the same procedure as \citet{breugem2005} we
shrink the kernel size as it approaches the top or bottom boundary, so that the
support of the kernel always remains inside the boundary. Owing to that, the
filtered fields fulfil the same boundary conditions at these boundaries as the
unfiltered fields (since the filtering kernel reduces to the Dirac delta
function). This generates the commutation error which is
proportional to the derivative of the filter size in space
\citep{sadowski2023}, therefore, without explicitly analysing these errors, the
momentum or scalar balance should not be analysed in these near-wall regions.

The velocity profile is clearly shifted towards the top wall, as a result of
different values of stresses between the top and bottom wall, as also reported
by \citep{kuwata2016}. The intrinsic velocity in the homogeneous region of the
porous medium (i.e., where gradients of average velocity and porosity are
negligible) is approximately equal to $U_c = 0.042U_\mathrm{bulk}$. This leads
to the Reynolds number based on the particle size $\mathrm{Re}_\mathrm{p} = U_c
	d_p / \nu = 11.7$, situating the flow in the porous region in the laminar drag
regime \citep{wood2020,demir2026}, which is typically characterized by a steady
state flow conditions and a potentially significant influence of inertia.

The fluctuations of velocity, with the root-mean-square defined as
\begin{equation}
	\psi_\mathrm{rms}= \sqrt{\langle{\overline{\psi'\psi'}}\rangle}
	\label{eq:rms}
\end{equation}
are
stronger near the porous wall than the top wall, and due to the relatively
permeable nature of the porous matrix, they slowly decay to zero with
decreasing $y/H$. In such a configuration, the porous matrix permits
large-scale velocity disturbances to pass through \citep{breugem2005,
	kuwata2024}. The rate of decay of the wall-normal fluctuations is noticeably
slower in comparison to $u^{t+}_\mathrm{rms}$ and $w^{t+}_\mathrm{rms}$, which
can be interpreted as the influence of Kelvin--Helmholtz stability which can
develop near the porous-fluid interface
\citep{breugem2005,kuwata2016,kuwata2024}.

The profiles of velocity and its fluctuations are aligned very well with the
data of \citet{breugem2005} and the results of \citet{chandesris2013}, which are
not plotted in the \cref{fig:velocity} for clarity. The
results of our simulation slightly overestimate the streamwise component of
velocity fluctuations.

\subsection{Influence of the filter on velocity and momentum closure}

\citet{chandesris2013} used \cref{eq:quadratic-kernel} to filter the results of
their DNS instead of \cref{eq:tent-kernel} as employed by \citet{breugem2005}
and demonstrated differences in computed values of macroscopic viscosity. They
also commented on inadequacy of the linear tent kernel to filter the
temperature fields. To explore the influence of the chosen kernel further, we
have decided to evaluate the filtered fields using three different filtering
molecules, \cref{eq:tent-kernel,eq:quadratic-kernel,eq:cubic-kernel}.

Overall, the choice of the filtering kernel has weak influence on the velocity
moments (results with different kernels plotted in \cref{fig:velocity}), with
differences only noticeable near top, bottom or porous wall --- larger
filtering molecule of the \cref{eq:quadratic-kernel,eq:cubic-kernel} leads to
smoothing of the peak of the stream wise velocity fluctuations. The tent filter
seems large enough to remove all microscale fluctuations from the velocity and
velocity fluctuations.

It is however not appropriate, as pointed out by \citet{chandesris2013}, to try
to determine the closure coefficients of the double-averaged fields, as they
demonstrated the adoption of \cref{eq:quadratic-kernel} allowed for limitation
of the fluctuations in the macroscopic turbulence viscosity, which for the current
configuration can be computed as
\begin{equation}
	\nu_{t\phi} = \frac{-\da{u'v'}}{\pdv{\da{u}}/{y}}.
	\label{eq:macro-viscosity}
\end{equation}

Macroscopic turbulent viscosity or effective viscosity is often used to model
the effect of turbulent dispersion \emph{via} the gradient hypothesis
\citep{antohe1997,lage2002} as indicated by \cref{eq:macro-viscosity}. More
sophisticated approaches can also be used, involving splitting the governing
equations into the contribution of macroscopic and microscopic scales
\citep{kuwata2014}, which involve two separate viscosities. The concept of
effective viscosity has also been recently employed by \citet{rao2025} to model
dispersion in configurations involving porosity gradient.

Overall, the distribution of $\nu_{t\phi}$ (see \cref{fig:velocity}) is the same as computed by
\citep{chandesris2013}. Near the bottom of the channel, the value of turbulent
viscosity is very small, due to negligible turbulent stresses and relatively
flat velocity profile. Here, the assumption
of neglecting $\tnsr{T}$ as typically done during the
derivation of averaged equations holds \citep{whitaker1996}. As the $y$
coordinate approaches the interface, both the velocity and macroscopic
turbulent shear stress increase, leading to positive values of macroscopic
viscosity. The use of the tent kernel leads to the oscillations of
$\nu_{t\phi}$ as observed by \citet{chandesris2013} and the use of quadratic
filtering function removes these macroscopic fluctuations, apart from the
changes in viscosity occurring near the interface. The use of higher-order
kernel does not remove this feature, nor it leads to further smoothing of
$\nu_{t\phi}$, suggesting that the changes occurring near the interface are not
an artefact of filtering procedure and an accurate macroscopic eddy-viscosity
model should reproduce them.

\begin{figure}
	\centering
	\subfigure[\label{fig:macro-vis}]{\includegraphics[width=8.5cm]{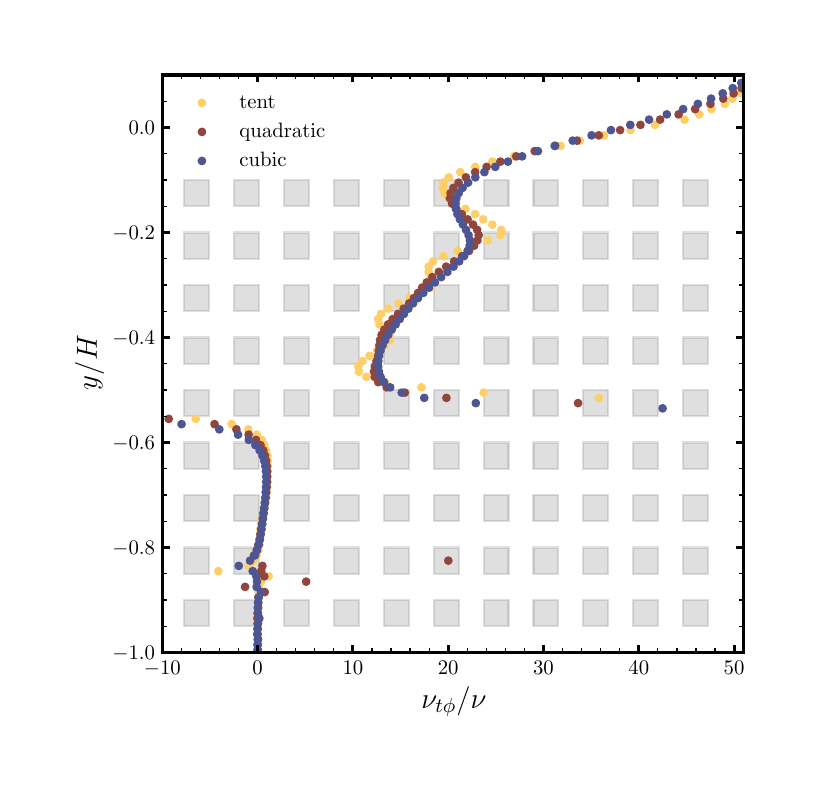}}

	\subfigure[the inverse of permeability ($1/\mathrm{Da}$, left), the
		Forchheimer number ($\mathrm{Fo}/\mathrm{Da}$, center) and the Forchheimer
		coefficient ($\mathrm{Fo}/(\mathrm{Da}\mathrm{Re}_\mathrm{p}$, right)
		\label{fig:drag}
	]{\includegraphics[width=8.5cm]{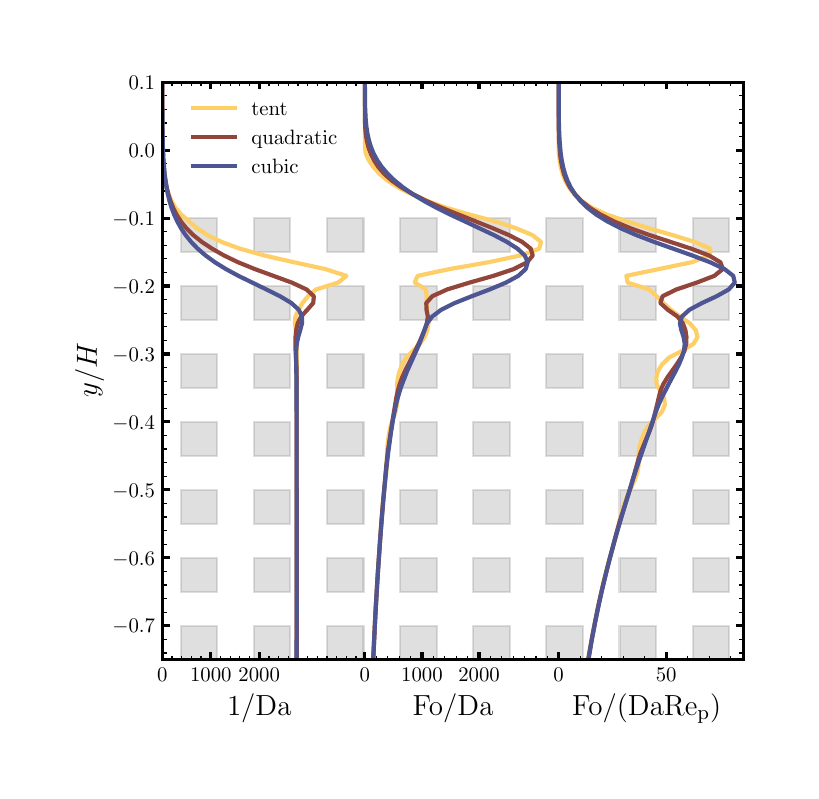}}
	\caption{Macroscopic turbulent viscosity (a) and modelling parameters of the
		drag closure (b) plotted near the interface and inside the porous
		medium. Results with three different filtering kernels are presented, tent
		kernel given by \cref{eq:tent-kernel}, quadratic kernel
		\cref{eq:quadratic-kernel} and cubic kernel
		\cref{eq:cubic-kernel}.}
\end{figure}

A stronger influence of the filtering kernel can be observed when inspecting
the profiles of closure parameters of porous-induced drag.
The drag can be modelled according to the Darcy--Forchheimer equation
\citep{whitaker1986,whitaker1996}, where a combination of two expressions,
linear and quadratic with respect to averaged velocity, is used to specify
drag. For the studied problem, this can be written as
\begin{equation}
	F_x = - \frac{H^2\phi^2\nu}{\mathrm{Da}}(1+\mathrm{Fo})\da{u}
	\label{eq:darcy-forch}
\end{equation}
where $\mathrm{Da}$ is the Darcy number representing permeability
($\mathrm{Da}= K/H^2$) and $\mathrm{Fo}$ is the Forchheimer number representing
influence of the inertial drag. The Forchheimer number is assumed to be
proportional to the flow velocity or $\mathrm{Fo}\propto\mathrm{Re}_\mathrm{p}$,
where $\mathrm{Re}_\mathrm{p} = \da{u}d_p/\nu$ is the particle size based Reynolds
number.

The profile of permeability for the present case can be computed from the
particle-resolved simulation conducted in the same domain under laminar flow
conditions \citep{sadowski2023c,sadowski2023b}, resulting in $\mathrm{Fo} = 0$.
Then \cref{eq:uncl-drag} can be evaluated directly and equalled to
\cref{eq:darcy-forch} to compute $K$. Using the computed value of permeability,
the Forchhaimer coefficient can than be determined from the current data using
the same procedure.

The resulting profiles of the inverse of permeability are plotted in
\cref{fig:drag} using each of the considered filtering kernels. Deep in the
homogeneous porous medium, permeability remains constant, however near the PFI,
permeability decreases before it increases rapidly. This behaviour was first
described by \citet{breugem2004} and was the focus of \citet{sadowski2023b},
where the choice of the filtering molecule and its size was shown to have a
strong impact on the shape of the permeability near the interface. The increase
of the filter size resulted in smoothing of this oscillation of permeability
and adopting of the Gaussian kernel allowed to remove it completely. In the
present case, adoption of the higher-order kernels
\cref{eq:quadratic-kernel,eq:cubic-kernel} corresponding to the cellular
average leads to the results consistent with the finding of
\citet{sadowski2023b}.

Moreover, as both linear and non-linear components of drag are large in
magnitude near the interface \citep{sadowski2023c}, the oscillations present in
the permeability profile directly impact the distribution of computed
$\mathrm{Fo}$, as exemplified in \cref{fig:drag}. Here, only the use of the
\cref{eq:cubic-kernel} leads to a smooth profile of the Forchheimer number,
with the oscillations present near the interface for both higher-order kernels
at coordinates aligned with the positions of the peaks of inverse permeability.
Interestingly, the use of the tent kernel also leads to small oscillations in
the homogeneous porous region. This is better visible if Forchheimer
coefficient $\mathrm{Fo}/\mathrm{Re}_\mathrm{p}$ is computed (right-most side
of \cref{fig:drag}), which is the closure coefficient determining the inertial
drag in the Darcy--Forchheimer model, when quadratic model for inertial drag is
assumed. The profile of $\mathrm{Fo}/\mathrm{Re}_\mathrm{p}$ is only smooth in
the homogeneous porous region for the cubic kernel, and below PFI, the use of
this kernel allows for minimization of the oscillations in the distribution of
the coefficient.
Another interesting feature of the computed profile of the
Forchheimer coefficient is that it is not constant in the porous medium, i.e.,
a drag model with quadratic inertial term, e.g., \citep{ergun1952} equation, will
not be able to properly represent the rise in $\mathrm{Fo}$ in this region.

A general observation follows, that the choice of the filtering
molecule when upscaling microscale results should be determined
based on the performed analysis. While linear kernel might be adequate for
inspecting first and second-order moments of velocity, a higher-order
cellular filtering is required for closure parameters undisturbed by microscale
fluctuations.

\subsection{Mean temperature}

\begin{figure}
	\begin{center}
		\includegraphics[width=8cm]{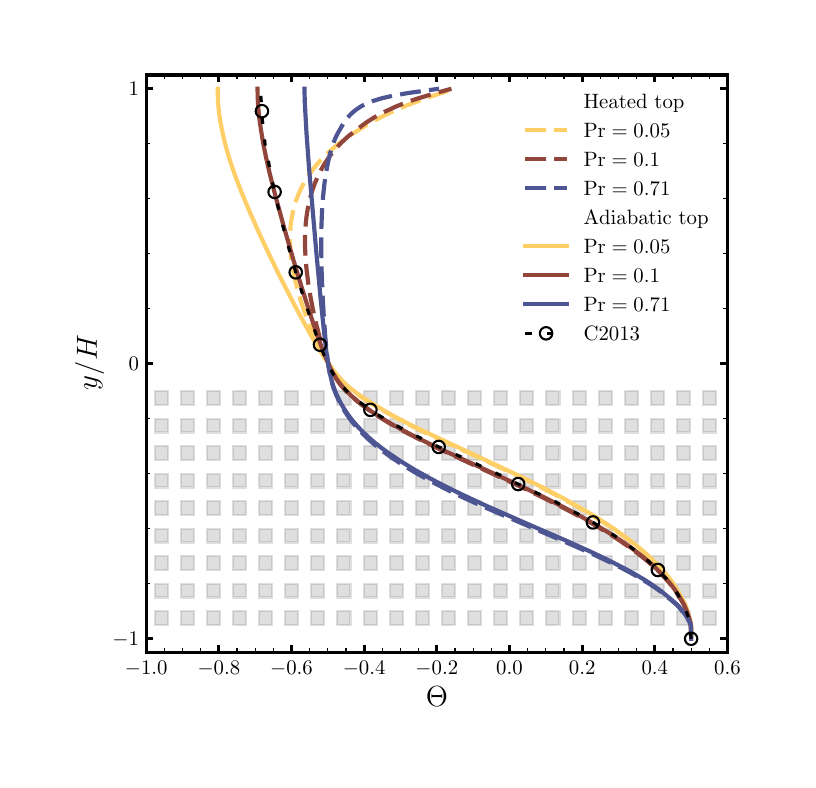}
	\end{center}
	\caption{Profiles of temperature normalized as given by \cref{eq:temp-norm}
		from cases with both boundary conditions, for each Prandtl number. The
		spatial filtering is performed using the quadratic kernel, i.e.,
		\cref{eq:quadratic-kernel}. The reference data corresponds to the results
		for $\mathrm{Pr}=0.1$ gathered by \citet{chandesris2013} for an adiabatic
		top wall configuration.}\label{fig:scalar-mean}
\end{figure}

The double-averaged temperature in the channel is plotted in
\cref{fig:scalar-mean}.
The temperature values are normalized as follows
\begin{equation}
	\Theta = \frac{
	\da{\theta} - (\overline{\theta}|_{y=0} + \overline{\theta}|_{y=-H})/2}{
	\overline{\theta}|_{y=-H} -\overline{\theta}|_{y=0} }
	\label{eq:temp-norm}
\end{equation}
resulting in double-averaged temperature spanning the range of $-0.5$ to $0.5$
inside the porous medium and allowing to easily compare the distribution of
$\da{\theta}$ in this region.
Following the observations of \citet{chandesris2013}
regarding the requirements for the filtering kernel in this configuration, we
use \cref{eq:quadratic-kernel}, i.e., the quadratic kernel, to compute the
spatially filtered fields. The use if tent kernel to filter the time-averaged
temperatures resulted in a presence of small scale fluctuations, which would
be, however, not noticeable in \cref{fig:scalar-mean}.

First and foremost, the result for the $q_t=0$ configuration and
$\mathrm{Pr}=0.1$ compares favourably with the DNS data gathered by
\citet{chandesris2013}, thereby, further validating the adopted simulation
approach. The change of the boundary condition at the top wall mainly
influences the temperature profile in the homogeneous fluid region, where a
thermal boundary layer forms adjacent the heated top wall and $\da{\theta}$
increases with the vertical coordinate. The temperature profile in the porous
region is not strongly affected by the boundary condition change, with small
deviation visible for the highest Prandtl number.

Above the cubes, the increase of $\mathrm{Pr}$ strengthens the influence of the
turbulence fluxes and flattens the temperature profile similarly to what can be observed in
turbulent flow over smooth walls \citep{kawamura1998}.
Due to low Peclet numbers inside the porous region ($\mathrm{Pe}_p =
	\mathrm{Pr}\mathrm{Re}_\mathrm{p} = 0.6, 1.2, 8.4$ for $\mathrm{Pr} = 0.05,
	0.1. 0.71$, respectively), the heat transfer in the region is
strongly influenced by the diffusive transport, and the temperature
distribution near the bottom wall of the channel resembles the parabolic
profile expected in fully laminar conditions in such geometry
\citep{chandesris2013}. The decrease of Prandtl number additionally results in
the steepening of the temperature gradient in the porous medium. This behaviour
is a direct result of the weakening of the turbulent heat transfer in the top
layers of the porous region.

\subsection{Turbulent heat flux}

\begin{figure*}
	\begin{center}
		\includegraphics[width=0.85\linewidth]{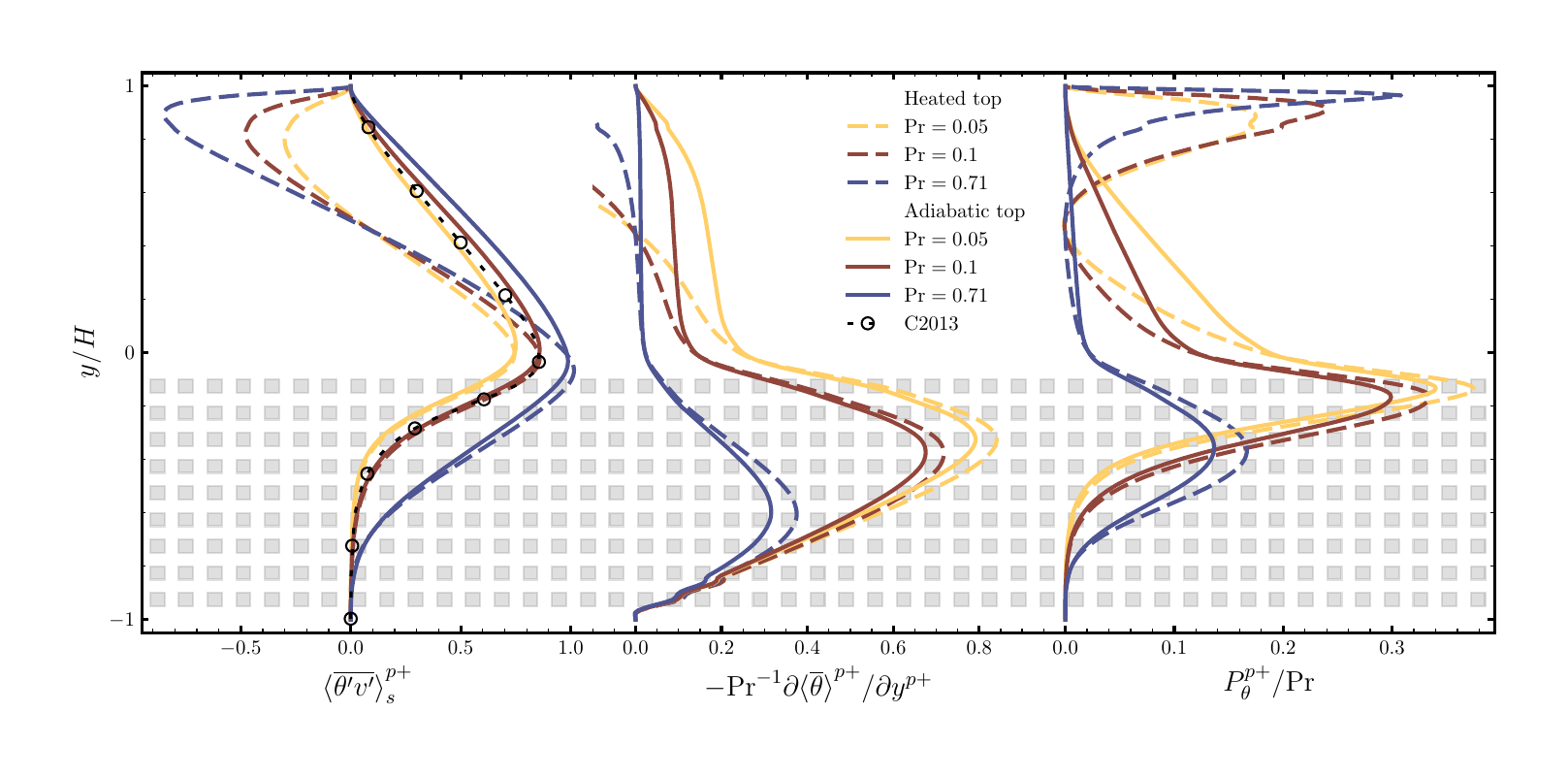}
	\end{center}
	\caption{The wall-normal turbulent heat flux in the channel (left), the gradient of mean temperature (centre) and the production of scalar variance (right) for both
		boundary condition configurations and computed Prandtl numbers. The spatial
		filtering is performed using the quadratic kernel, i.e.,
		\cref{eq:quadratic-kernel}. The reference data corresponds to the results
		for $\mathrm{Pr}=0.1$ gathered by \citet{chandesris2013} for an adiabatic top
		wall configuration.
	}\label{fig:v-turb-heat-flux}
\end{figure*}

The profiles of turbulent heat flux ${\da{\theta'v'}}_s$ are plotted in
\cref{fig:v-turb-heat-flux} along with the data gathered by
\citet{chandesris2013}. Computed results seem to overestimate slightly the
values of the heat flux in the clear channel part of the domain, however the
overall agreement between the profiles for $\Pr{}=0.1$ is very good. The
difference might be stemming from the different methods of computation of
$\theta_\tau^p$ --- \citet{chandesris2013} followed \citet{breugem2005} and
used the balance of filtered quantities instead of the global energy balance,
which leads to a small difference in $\theta_\tau^p$ analogously to the friction
velocity.

The change of boundary conditions does not influence the profile of the
turbulent heat flux in the porous region, with only small deviation visible for
the highest Prandtl number case. Overall, the increase of $\Pr{}$ allows for
deeper penetration of the turbulent fluxes into the porous medium, with
${\da{\theta'v'}}_s$ vanishing for approximately $y/H\leq0.9, 0.8$ and $0.65$
for $\Pr = 0.71, 0.1$ and $0.05$, respectively. Above that, the turbulent flux
adopts nearly linear profile, with the slope increasing as $\Pr{}$ is reduced.
The weak dampening of the turbulent heat flux inside the porous medium can be
mainly attributed to the permeability of the studied porous structure: the results of
\citet{nishiyama2020a} indicate that high streamwise permeability especially leads to higher values of
turbulent heat flux.
The maximum of each profile is located above the top-most layer of cubes, near
the position of the peak of wall-normal fluctuations.

\begin{figure}
	\begin{center}
		\includegraphics[width=8cm]{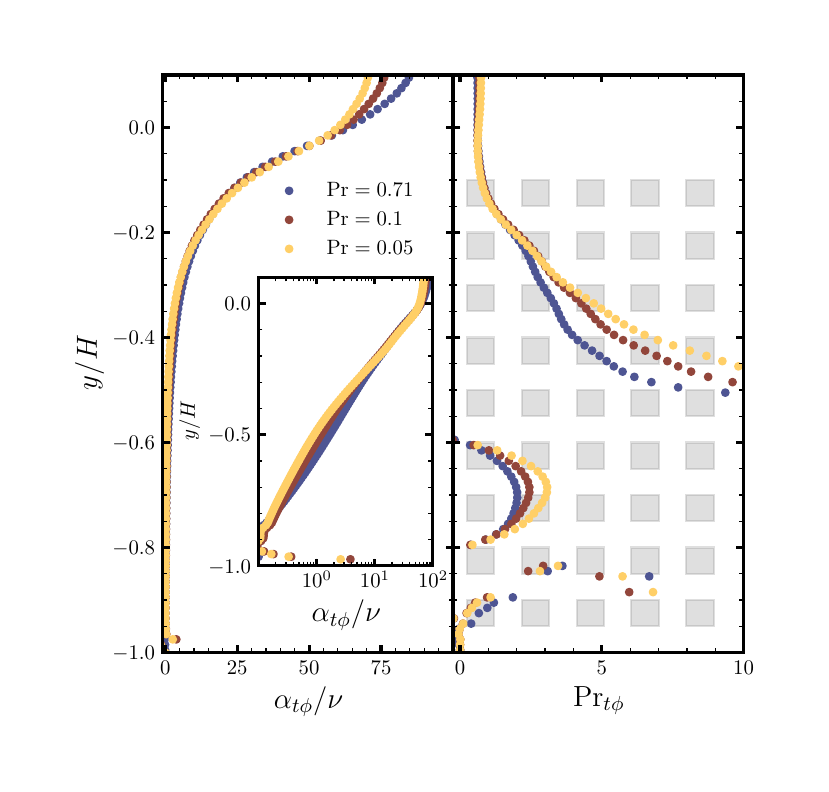}
	\end{center}
	\caption{The computed values of the macroscopic turbulent heat transfer
		coefficient $\alpha_{t\phi}$ (left) and corresponding turbulent Prandtl
		number $\mathrm{Pr}_{t\phi}$ (right).}\label{fig:macro-alpha}
\end{figure}

Similarity, to the turbulent momentum flux, the macroscopic turbulent heat flux
can be modelled via the gradient hypothesis, where the turbulent diffusivity
can be introduced
\begin{equation}
	\alpha_{t\phi} = \frac{-\da{\theta'v'}}{\pdv{\da{\theta}}/{y}}.
	\label{eq:turb-heat-transfer-coeff}
\end{equation}
The distributions of $\alpha_{t\phi}$ are plotted in \cref{fig:macro-alpha}
corresponding to the three Prandtl numbers for the case with adiabatic
top-wall. The gathered data for $\Pr{}=0.1$ is in good agreement with the
results evaluated by \citet{chandesris2013} (the profile not plotted in
\cref{fig:macro-alpha} for clarity).
The distributions of $\alpha_{t\phi}$ remain very similar across the Prandtl numbers,
which is surprising considering the strong influence of $\mathrm{Pr}$ on both the
distributions of $\da{\theta'v'}$ and the gradient of $\da{\theta}$, plotted in \cref{fig:v-turb-heat-flux}.

The turbulent heat flux is weak near the bottom wall, while in the same region,
the gradient of temperature is already
decreasing (the osscilations of the temperature gradients near the top wall are
a result of decreasing the filter size as $y$ approaches the bottom wall).
Generally, the heat flux starts to rise only around the vertical coordinate
corresponding to the maximum of the profile of temperature gradient. Hence, the
values of diffusivity remain quite low until $y/H\approx 0.4$, above which the
decrease of the temperature gradient combined with nearly linear growth of
$\da{\theta'v'}$ results in sharp increase of the diffusivity. The whole
distribution of $\alpha_{t\phi}$ in the porous medium follows exponential
relation as can be identified from the linear distribution of $\alpha_{t\phi}$
in the inset plot in \cref{fig:macro-alpha} with a logarithmic scale.

The macroscopic turbulent Prandtl number $\mathrm{Pr}_{t\phi} =
	\nu_{t\phi}/\alpha_{t\phi}$ is plotted in \cref{fig:macro-alpha} as well. It
can be split into three regions in the porous medium, divided by the singular
points in the profile of $\nu_{t\phi}$. Near the bottom wall, $\Pr{}_{t\phi}$
is increasing with $y$, however, the scatter of the values is quite large here.
The distribution changes sign around $y/H=0.8$, and for $ -0.8 < y/H < -0.6$,
the differences between the three Prandtl numbers start to be noticeable.

The small
differences in $\alpha_{t\phi}$ gets attenuated leading to $\Pr_{t\phi}$ being
inversely proportional to $\Pr{}$, with the largest maximum
$\mathrm{Pr}_{t\phi}\approx4$ corresponding to $\Pr=0.05$. Above $y/H \approx
	0.6$, the value of $\nu_{t\phi}$ changes abruptly, leading also to a jump of
$\Pr_{t\phi}$ which decreases with as $y$ comes closer to PFI. Here,
initially the spread of the three $\mathrm{Pr}$ distributions is large, with
the same trend as observed below (highest $\Pr_{t\phi}$ for lowest
$\mathrm{Pr}$). With increasing vertical coordinate the values of
$\Pr_{t\phi}$ decrease from $\Pr_{t\phi}>10$ to $\Pr_{t\phi}\approx0.7$
for each of the considered Prandtl numbers, which is in range
of typically assumed values \citep{sagaut2006}.
The differences between profiles of $\Pr_{t\phi}$ also decreases with $y$: near
the interface, at $y/H\approx0.25$ the distributions overlap.
The profile of $\alpha_{t\phi}$ is much less sensitive to the choice
of the filtering molecule than $\nu_{t\phi}$. Although, the oscillations stemming
from inadequate filtering molecule can be observed in temperature gradients (as
discussed in \cref{sec:scalar-balance}), since the magnitude of the gradient of $\da{\theta}$
is large in most of the porous medium these oscillations do not result in large changes
in $\alpha_{t\phi}$.

\subsection{Temperature fluctuations}
\begin{figure}
	\begin{center}
		\includegraphics[width=8cm]{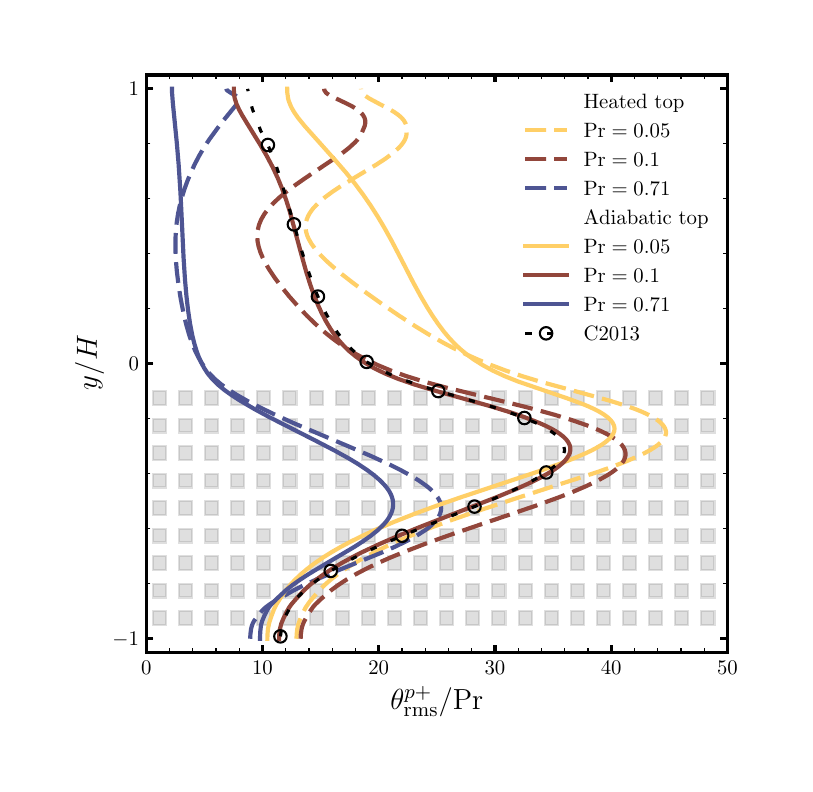}
	\end{center}
	\caption{The fluctuations of the temperature fields in the channel for both
		boundary condition configurations and computed Prandtl number. The spatial
		filtering is performed using the quadratic kernel, i.e.,
		\cref{eq:quadratic-kernel}. The reference data corresponds to the results
		for $\mathrm{Pr}=0.1$ gathered by \citet{chandesris2013} for an adiabatic top
		wall configuration.}\label{fig:scalar-rms}
\end{figure}

The profiles of the fluctuations of temperature field $\theta_\mathrm{rms}$,
defined using \cref{eq:rms}, are plotted in \cref{fig:scalar-rms}. As expected,
the increase of Prandtl number leads to stronger temperature fluctuations.
Moreover, a good agreement between the current results and the data of
\citet{chandesris2013} can be observed. The distributions of
$\theta_\mathrm{rms}$ are not sensitive to the choice of the filtering kernel,
similarly to the velocity variances.

For each of the studied cases, the maximum of $\theta_\mathrm{rms}$ is
positioned inside the porous medium. For the case with the heated top wall, the
variance of temperature is still far larger inside the porous medium than in
the boundary layer near the top wall. Contrary to the mean temperature profile,
the value of heat flux at the top wall has a strong influence on the profile of
$\theta_\mathrm{rms}$. The presence of the heated wall results in a peak
in turbulent fluctuations in the boundary layer adjacent to the top wall, which
do not vanish at the wall as found by \citet{tiselj2001}. Moreover,
the variance in the middle of the channel is weaker for the heated top wall.

For each $\mathrm{Pr}$, the character of the profile of the $\theta_\mathrm{rms}$
remains similar inside the porous medium for both investigated boundary
condition configurations. However, for the heated wall case, the turbulent
fluctuations inside a porous medium are stronger.

\citet{chandesris2013} had reasoned that the increase of
$\theta_\mathrm{rms}$ inside the porous medium is induced by large scale
turbulent fluctuations caused by pressure waves, penetrating the porous medium
and increasing the variance of the temperature. \citet{kuwata2020} also
reported on the same effect in their simulation of heat transfer in a duct with
a porous layer. The work of \citet{nishiyama2020a}, considering similar channel
flow configuration, focused on this issue further. Based on the distribution of
the production term of the temperature variance $P^\theta$ \citep{zhou2019}:
\begin{equation}
	P_\theta = -\da{v'\theta'}\pdv{\da{\theta}}{y},
	\label{eq:production}
\end{equation}
the rapid increase of the fluctuations inside the porous medium has been linked
to the interaction between the strong turbulent heat flux induced by the large
scale unsteady motions and the gradient of the mean temperature.

The profiles of the temperature gradient and production are plotted together
with $\da{v'\theta'}$ in \cref{fig:v-turb-heat-flux}. For lower Prandtl
numbers, the peak of $-\pdv{\da{\theta}}/{y}$ is shifted closer to the
interface. Meanwhile, as previously discussed, the penetration depth of the
turbulent heat flux decreases with $\mathrm{Pr}$. Both these effects together
result in the shift of the overall distribution of $P_\theta$, together with
$\theta_\mathrm{rms}$, upwards as $\mathrm{Pr}$ decreases.
The strengthening of
turbulent fluctuations inside the porous medium in the heated top wall configuration, can
also be explained by analysing the distribution of production. Although, the
turbulent heat flux is not strongly affected by the change of boundary conditions, the
temperature distribution and gradient inside the porous medium are, leading to a stronger
production of temperature variance in the porous region.

\subsection{Temperature balance in the channel}
\label{sec:scalar-balance}

\begin{figure*}
	\begin{center}
		\includegraphics[width=0.95\textwidth]{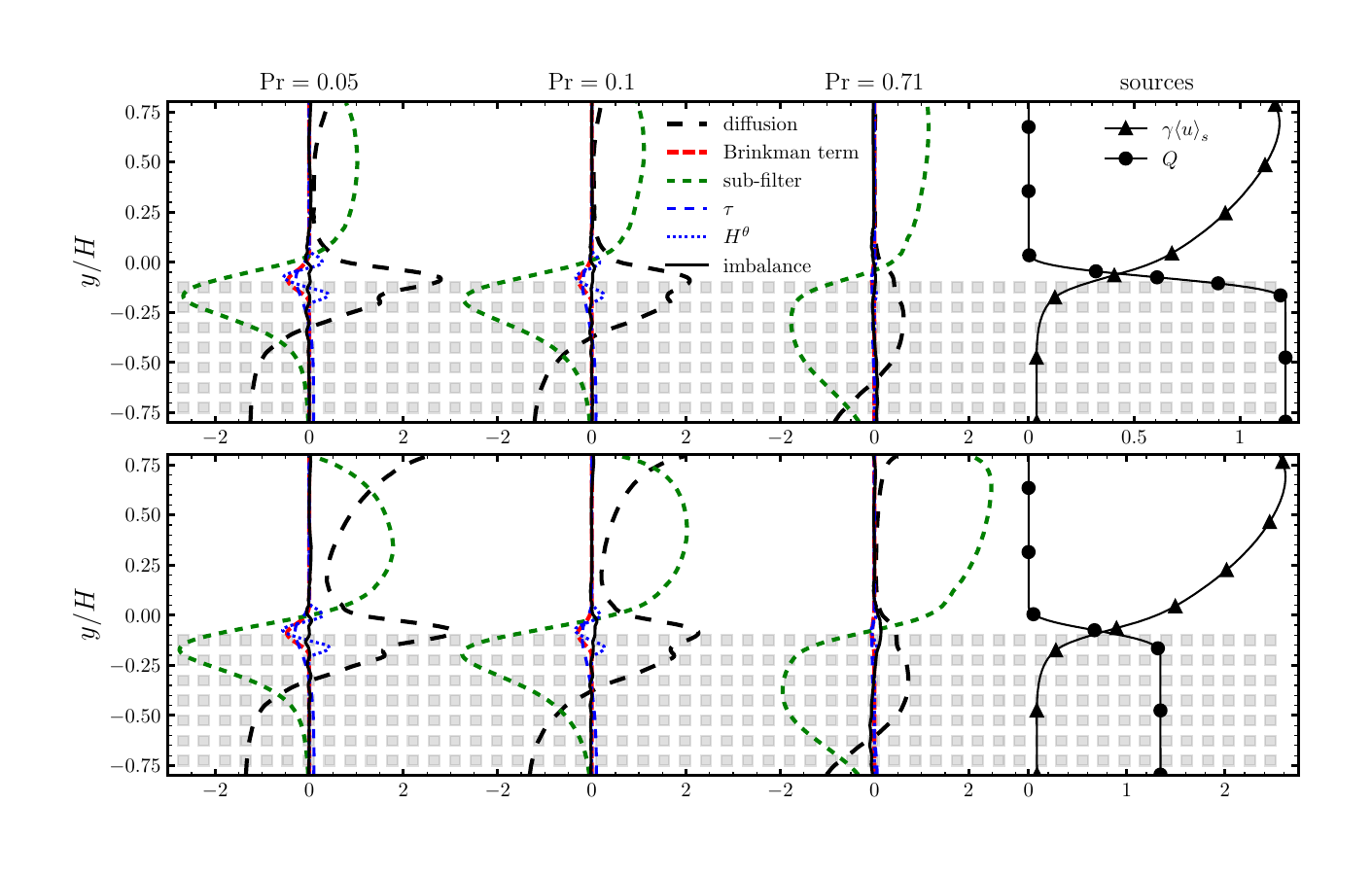}
	\end{center}
	\caption{The temperature budget terms computed for each Prandtl number (columns),
		for the adiabatic top wall configuration (top row) and the heated top-wall
		configuration. The terms were evaluated using the quadratic kernel given by \cref{eq:quadratic-kernel}.}
	\label{fig:scal-balance}
\end{figure*}

Considering that the studied double-averaged flow is statistically
one-dimensional, the temperature balance in the channel can be written as
\begin{equation}
	\begin{split}
		\underbrace{\phi\da{u}\gamma - Q}_{\text{sources}} =
		\underbrace{\phi\alpha\pdv{\da{\theta}}{y,y}}_{\text{diffusion}}
		+ \underbrace{\alpha\pdv{\phi}{y}\pdv{\da{\theta}}{y}}_\text{Brinkman term}
		+ \tau \\
		+ \underbrace{\pdv{}{y} \ab(\phi\da{v}\da{\theta} - \phi\da{v\theta}) }_\text{sub-filter stress}
		+ \mathcal{H}^{\theta}.
	\end{split}
	\label{eq:balance}
\end{equation}

The \cref{fig:scal-balance} presents the temperature balance computed for each
Prandtl number in the whole channel. The comparison of current results to DNS
velocity and scalar data of \citet{breugem2005, chandesris2013} suggest that
a comparable accuracy has been achieved in current results. Both implicit and
explicit influence of LES modelling does not alter the gathered statistics
significantly and adopted mesh resolution is indeed sufficient. Hence, the term
encompassing the explicit influence of the LES model has been omitted from the
above balance equation, and can be understood to be lumped into the plotted
imbalance of \cref{eq:balance}.

Moreover, the imbalance is an additional measure of the convergence of statistics.
The imbalance is small for each of the cases
allowing a conclusion that adopted averaging time are overall sufficient for
the present analysis. The biggest values of imbalance are clustered near the
interface, for the $\mathrm{Pr}=0.71$ case with the heated top wall. The
separation of time-scales is the largest for this case as discussed in
\cref{sec:Geometryflow}, and the boundary condition configuration with the heat flux
prescribed at the top wall was averaged over a shorter time than the adiabatic
top wall case. That said, the magnitude of the error is still smaller than the
rest of the significant terms comprising the budget for this Prandtl number,
allowing for the analysis.

The boundary condition configuration does not impact the distribution of terms
in the temperature balance inside the porous medium. The only changes occur in
the homogeneous fluid region, where the presence of the thermal boundary layer
near the top wall results in different profiles of molecular and turbulent
diffusion (i.e., the divergence of sub-filter stresses; the contribution of
dispersion due to filtering is negligible).

The bottom-most layers of the porous medium are dominated by molecular
diffusion. This region generally shrinks with the Prandtl number as the
influence of turbulence transfer on the budget becomes more pronounced, due to
the increased strength and penetration depth of the turbulent fluxes. Above,
both molecular and turbulent diffusion, are significant.
The comparative strengths of molecular and turbulent diffusion inside the
porous medium are in line with what can be estimated based on the Peclet
numbers. For lower $\Pr$, $\mathrm{Pe}_\mathrm{p}\sim 1$ and the magnitude of
both diffusions are similar, whereas for the highest Prandtl number
($\mathrm{Pe}_\mathrm{p}=8.4$) the turbulent diffusion becomes around 3 times
stronger.

The character of the diffusion profile changes significantly between the two
lowest Prandtl numbers and the $\Pr=0.71$. As $\Pr$ decreases, the increased
gradient of temperature at the PFI leads to a strong peak of the molecular
diffusion, positioned above the top layer of cubes. This feature is almost not
visible for the highest $\Pr$ cases, with the maximum of diffusion position
inside the porous medium. As the distribution of ${\da{v'\theta'}}_s$ becomes
more linear in the porous medium with the increase of $\Pr{}$ (see
\cref{fig:v-turb-heat-flux} and related discussion) the profile of sub-filter
term in the balance flattens, however, the peak of turbulent diffusion is
always positioned inside the porous medium.

The inspection of the Brinkman term, tortuosity and $\mathcal{H}^\theta$ (which
include the terms up to second-order) reveals that, for the present geometry
and boundary conditions, these terms can be fully neglected for the highest
$\Pr$, as their contributions are much smaller than the other terms. For lower
$\Pr$ all three increase in value. The tortuosity is small in magnitude and
positive in the whole porous medium. Near the interface, it becomes negative
and increases it impact on the budget (the tortuosity will be discussed in
detail in \cref{sec:hot}). The values of the Brinkman term tortusity and,
surprisingly $\mathcal{H}^\theta$, are comparable near the interface for lower
$\Pr$, indicating that each of these terms would require explicit modelling for
an accurate description of interface flows of low Prandtl number fluids.
Moreover, representing $\mathcal{H}^\theta$ up to second order is necessary to
reduce the error in the temperature balance.

\begin{figure}
	\begin{center}
		\includegraphics[width=8.5cm]{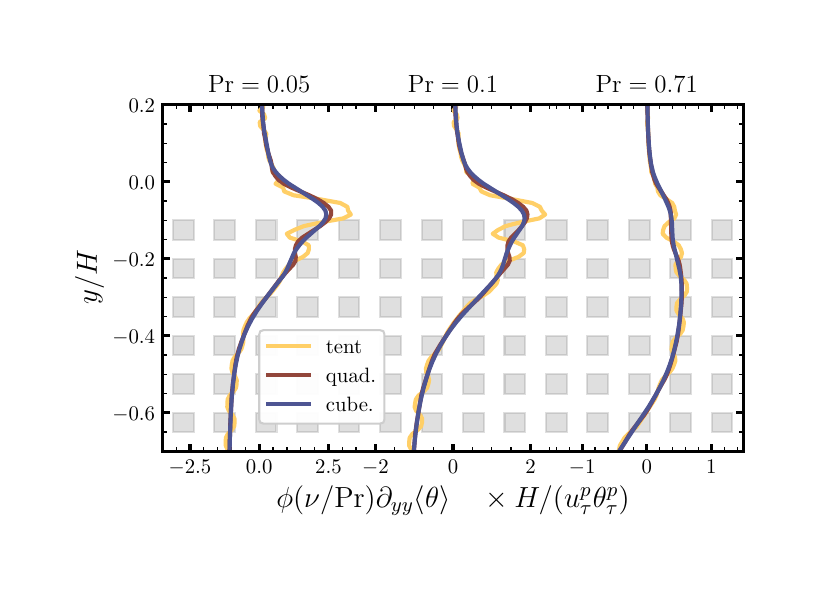}
	\end{center}
	\caption{Molecular diffusion term of the temperature budget plotted for each
		Prandtl number, using all three different filtering kernels
		(\cref{eq:tent-kernel,eq:quadratic-kernel,eq:cubic-kernel}, i.e., linear,
		quadratic and cubic functions respectively) and are normalised using the friction temperature, velocity and the
		height of the channel.}\label{fig:diffusion-kernels}
\end{figure}

Importantly, the evaluation of the temperature balance becomes sensitive to the
choice of the filtering kernel \emph{via} the distribution of the molecular
diffusion, which is plotted for each of the kernels in
\cref{fig:diffusion-kernels}. For the tent kernel, diffusion is polluted by
microscale fluctuations, clearly related to the presence of the cubes, but the
strength of these fluctuations weakens with the Prandtl number. Only the cubic
kernel smoother out the oscillation in diffusion positioned below the top-most
layer of cubes, however, overall, the cubic kernel is sufficient for the
removal of most fluctuations and the analysis of the terms in the temperature
balance.

\subsection{Tortuosity flux}
\label{sec:hot}

\begin{figure*}
	\begin{center}
		\includegraphics[width=0.8\textwidth]{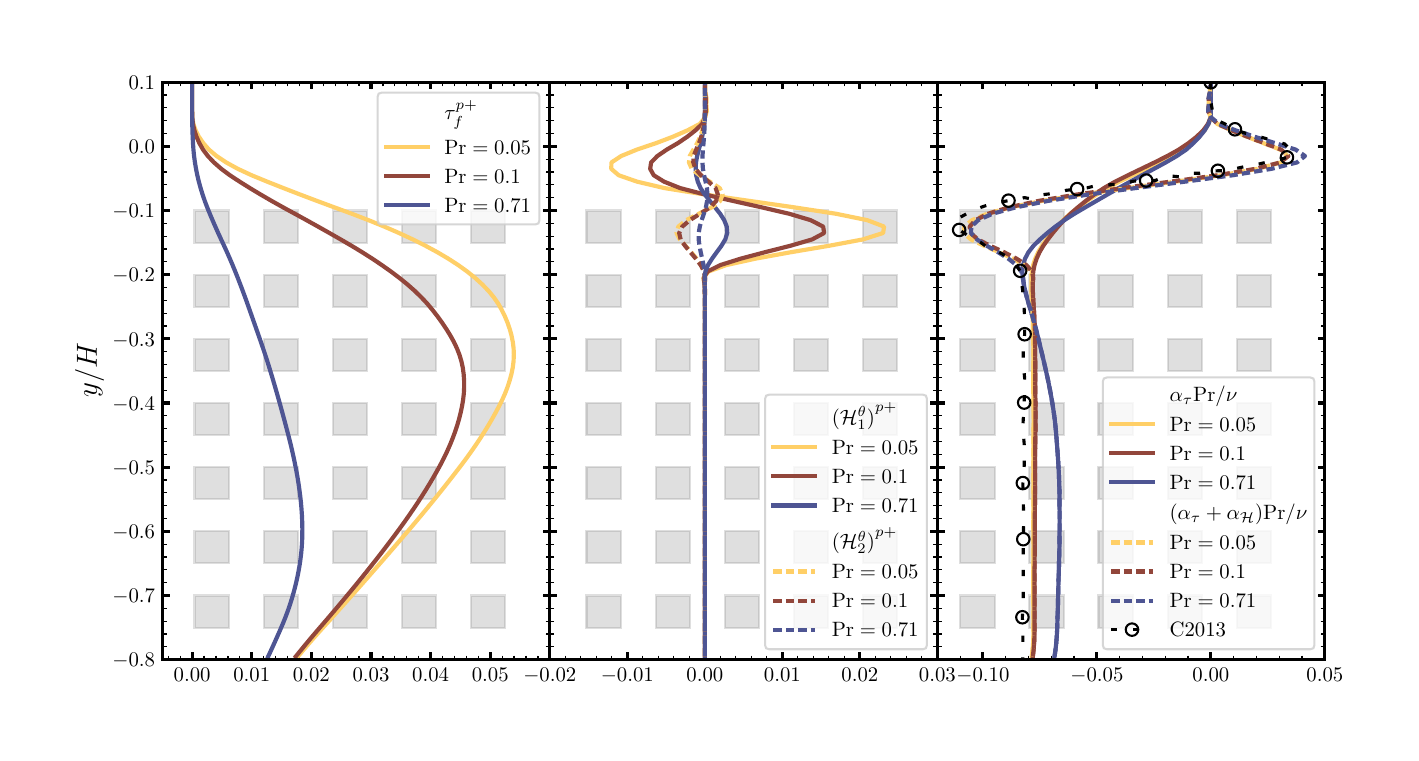}
	\end{center}
	\caption{The profiles of the tortuosity flux (left) the terms originating
		from the Taylor expansion of the double-averaged values inside the surface
		integral (centre) and the related effective viscosities (right), computed using
		the quadratic filtering kernel.}\label{fig:tortuosity}
\end{figure*}

Lastly, we analyse the tortuosity $\tau$ and the related term $\mathcal{H}^\theta$.
The flux of tortuosity, which we denote $\tau_f$ is typically modelled using
an effective viscosity model
\begin{equation}
	\tau_f = \int\limits_A G \alpha\va*{\overline{\theta}} n_i\odif{A} = \alpha_\tau \pdv{\da{\theta}}{x_i},
	\quad \tau = \pdv{}{x_i}\ab(\alpha_\tau\pdv{\da{\theta}}{x_i})
	\label{eq:tort-flux}
\end{equation}
and, in the case of current geometry, $\alpha_\tau = \tau_f /
	(\pdv{\da{\theta}}/{y})$. The distributions, of both $\tau_f$ and $\alpha_\tau$
are plotted in \cref{fig:tortuosity} from the case with the adiabatic top wall
(as established by \citet{chandesris2013} and in the previous section, the
boundary conditions do not influence the profiles of tortuosity).

Overall, the shape of $\tau_f$ bears strong resemblance to the gradient of
temperature (see \cref{fig:v-turb-heat-flux}), which motivates the effective
viscosity assumption. The resulting profile of $\alpha_\tau$ is nearly constant
in the porous medium for two lower $\Pr$, where it adopts a negative value. At
the interface, the effective viscosity smoothly transitions to 0. For the
highest value of $\Pr$, a similar behaviour is observed, but the value of
$\alpha_\tau$ is not constant --- it decreases slowly as the vertical
coordinate moves away from the interface. Possibly, this situation is related
to much stronger influence of the turbulent fluxes in porous region at
$\Pr=0.71$ as indicated by the budget analysis.

Importantly, the profile of $\alpha_\tau$ is different that the one obtained by
\citet{chandesris2013}, i.e., the oscillations near the interface region are
missing.
For the one-dimensional temperature balance, the terms resulting from the
Taylor series expansion during the derivation of averaged equations can be
written as
\begin{equation}
	\mathcal{H}^\theta = -\pdv{}{y}\Bigg(
	\underbrace{\alpha \pdv{ \va_{\xi_y}}{y}\pdv{\da{\theta}}{y}}_{\mathcal{H}^\theta_1}
	+ \underbrace{\alpha \pdv{\xi_y^2}{y}\pdv{\da{\theta}}{y,y}}_{\mathcal{H}_2^\theta}
	\Bigg),
	\label{eq:hot-1d}
\end{equation}
up to second-order.
This results of \citet{chandesris2013} can be recovered if the fluxes $\mathcal{H}^\theta_1$ and
$\mathcal{H}^\theta_2$, defined as in \cref{eq:hot-1d}, are included in the
tortusity flux as $\tau_f - \mathcal{H}_1^\theta - \mathcal{H}_2^\theta$,
forming a total effective viscosity $\alpha_\tau+\alpha_\mathcal{H}$
plotted in \cref{fig:tortuosity}.

Alternatively, one can compute the
\emph{exact} tortuosity flux, termed here $\tau^*_f$:
\begin{equation}
	\tau^*_f =  \da{\theta}\pdv{\phi}{x_i} + \int\limits_A G \alpha\overline{\theta} n_i\odif{A}.
	\label{eq:tort-direct}
\end{equation}
In the above, directly using the time-averaged variables instead of
fluctuations from the spatial mean includes all the terms resulting from the
Taylor series expansion (see \cref{eq:tort}).

Both first and second-order fluxes, $\mathcal{H}_1^\theta$ and
$\mathcal{H}_2^\theta$ computed using the quadratic kernel, are shown in
\cref{fig:tortuosity}.
The first-order flux is much stronger than $\mathcal{H}_2^\theta$ and coincidentally, it is also easier to be represented with the effective viscosity.
The $\pdv{{\va{\xi_\alpha}}_s}/{x_i}$ can be taken as additional tensorial
viscosity, that is determined purely from the geometry. The importance of this
term was already studied by \citet{battiato2019} in a scalar transport case in
a periodic array of squares, where a mollified tent function was used for
filtering. Although the inclusion of the related viscosity improved the
prediction \emph{vs.} the filtered DNS, the influence of $\mathcal{H}_1^\theta$
was much weaker than presented here, underlying the importance of these
terms at the interface, for low Prandtl numbers.

\begin{figure}
	\begin{center}
		\includegraphics[width=8.5cm]{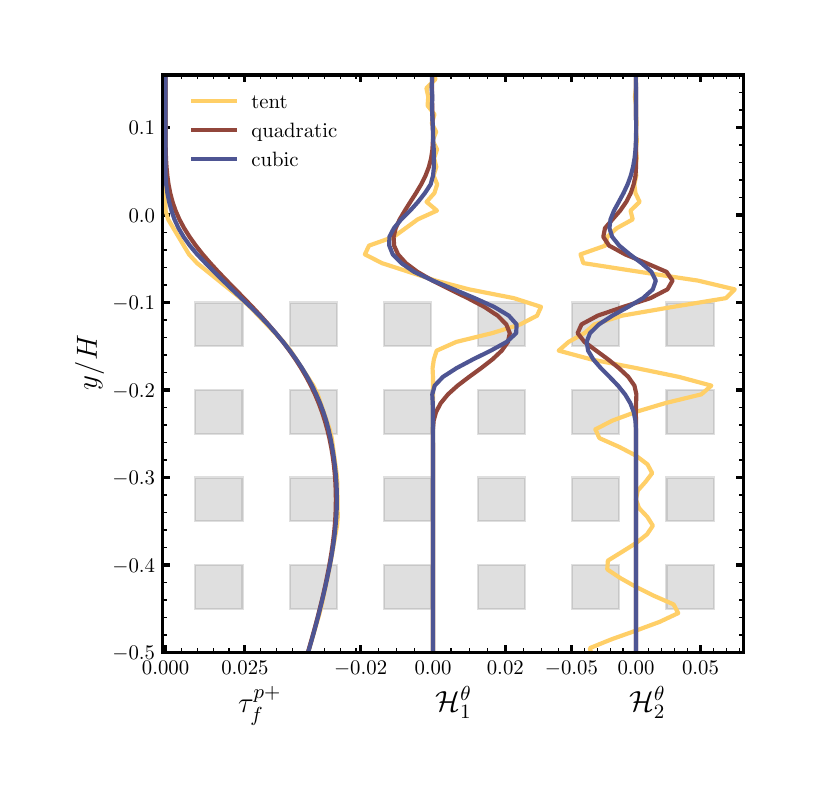}
	\end{center}
	\caption{The profiles of the tortuosity flux (left) the terms originating
		from the Taylor expansion of the double-averaged values inside the surface
		integral: first-order (centre) and the second-order (right), computed using
		tent, quadratic and cubic kernels, i.e.,
		\cref{eq:tent-kernel,eq:quadratic-kernel,eq:cubic-kernel}.}\label{fig:tortuosity-kernel}
\end{figure}

Both higher-order fluxes are only active in the interface region. Inside porous
medium ${\va{\xi_y}}_s$ and ${\va{\xi^2_y}}_s$ are constant for the chosen
filtering molecule, i.e., \cref{eq:VAM:geometric-ordered} is satisfied.
This can be observed in the \cref{fig:tortuosity-kernel}, where $\tau_f$ and
both $\mathcal{H}_1^\theta$ and $\mathcal{H}_2^\theta$ are plotted using each
of the considered kernel. In case of the tortuosity flux, the influence of the
filtering molecule is barely perceptible, with the increased order of the
cellular average smoothing the distribution of $\tau_f$ near the PFI. The
first-order term is strongly impacted --- a narrower support of the tent
function results in sharper gradients at the interface which increases
magnitude of $\mathcal{H}^\theta_1$. It is also clear, that the linear kernel
is sufficient to ensure that the first moment of the kernel ${\va{\xi_y}}_s$
vanishes in the homogeneous porous medium. It does not however, ensure the
vanishing of the second moment ${\va{\xi^2_y}}_s$, resulting in the
$\mathcal{H}^\theta_2$ being polluted by oscillations inside the porous medium,
which are stronger in magnitude than the values computed using the higher-order
kernels. In case of both high-order fluxes, the change from the quadratic to
the cubic kernel does not result in significant shift in plotted profiles.

\section{Conclusions}
\label{sec:Conclusions}

The turbulent flow ($\Re_\mathrm{bulk}=5485$) with heat transfer in a channel half-filled with an array
cubes with porosity of $0.875$ has been simulated \citep[the geometry of][]{breugem2005}. The
temperature field has been represented as a passive scalar and results have been
gathered for three different Prandtl numbers: $\Pr=0.71$, $0.1$ and $0.05$. The
time-averaged particle-resolved solution has been filtered using three
different filtering kernels (tent, i.e., linear, quadratic and cubic), each
constructing so-called cellular average.

The increase of Prandtl number leads to a decrease of the temperature
gradient inside the porous medium and a deeper penetration of turbulent fluxes.
The macroscopic turbulent viscosity, modelling the turbulent heat flux \emph{via}
gradient hypothesis, is weakly influenced by the Prandtl number. Small changes
between Prandtl numbers are visible in the distribution of turbulent macroscopic
Prandtl number, with the highest values computed for $\Pr=0.05$ case.

The maximum of the temperature variance is positioned inside the
porous medium, with the increase of $\Pr$ shifting the position of the
peak closer to the bottom wall. As proposed by
\citet{nishiyama2020a}, this behaviour has been linked to the production of
temperature variance. The profile of production is also shifted
downwards with $\Pr$, which is caused by the interplay between the
turbulent heat flux and mean temperature gradient.

At the highest $\Pr=0.71$ the contribution of Brinkman term, tortuosity
and the higher order terms of the Taylor series of filtered variables
(which are typically neglected during the deviation of averaged equations)
has been found to not contribute significantly to the temperature budget.
At lower Prandtl number, these terms are all comparable, and significant at
the porous-fluid interface. Therefore, especially while modelling the heat
transfer at low $\Pr$, they should not be neglected.

The temperature profile and profiles of the terms constituting the
temperature balance are only weakly impacted inside the porous medium by
the change of the boundary condition near the top wall, however, a slight change of
the mean temperature gradient leads to a stronger production of variance and, hence, stronger
fluctuations of temperature fields inside the porous medium.

The inclusion of the expressions containing the first and second-order terms of
the Taylor expansion of mean temperature, has been found to be necessary to
close the momentum balance. Their omission during the derivation would lead to
errors in the predicted temperature, but their significance at the interface is
much grater than what has been shown in the work of \citet{battiato2019}, who
evaluated these terms up to first order in homogenous porous medium. An
adequate choice of the filtering kernel is crucial to evaluate these terms as an
insufficient kernel order will lead to non-zero kernel moments in the
homogenous porous medium.

Distributions of the first and second-order statistics of the velocity and the
temperature are generally not impacted by the choice of the filtering molecule.
The amplitude of microscale oscillations remaining in the macroscopic fields
after filtering is insignificant for the tent kernel for both temperature and
velocity. On the other hand, the computation of closure parameters is typically
strongly impacted: fluctuations related to the geometry of the porous medium
are introduced by computing gradients of mean properties or dividing by a small
filtered quantity. This is best exemplified by computing the porous induced
drag, where differences in permeability profile due to the chosen kernel
influence the determination of Forchheimer coefficient as well. Hence, the use
of higher order kernels is advised for determination of the unclosed terms, but
tent kernel is sufficient when only first or second moments are of interest.

\printcredits

\nocite{}
\bibliographystyle{cas-model2-names}

\bibliography{references}

\end{document}